\documentstyle[11pt]{article}
\begin{document}
\normalbaselineskip=16 true pt
\normalbaselines
\bibliographystyle{unsrt}

\def\cone {\chi_1^0}
\def\ctwo {\chi_2^0}
\def\mc {m_{\chi_1^0}}
\def\L {\Lambda}
\def\s {\sigma}
\def\grav {\tilde G}

\def\be {\begin{equation}}
\def\bea {\begin{eqnarray}}
\def\ee {\end{equation}}
\def\eea {\end{eqnarray}}

\def\b {\bibitem}

\def\r {\rightarrow}
\def\lr {\longrightarrow}


\begin{flushright}
MRI-PHY/P970615\\
hep-ph/9707239
\end{flushright}

\vskip 1 true cm

\begin{center}

{\Large\bf
{TESTING GAUGE-GRAVITINO COUPLING IN
GAUGE-MEDIATED SUPERSYMMETRY BREAKING
THROUGH SINGLE PHOTON EVENTS}}\\[10mm]
{\large\em Amitava Datta$^{*}$\footnote{Electronic address: 
adatta@juphys.ernet.in}, 
Aseshkrishna Datta$^{*}$\footnote{Electronic address: 
asesh@juphys.ernet.in}, 
Anirban Kundu$^{**}$\footnote{Electronic address: akundu@mri.ernet.in},\\
Biswarup Mukhopadhyaya$^{**}$\footnote{Electronic address: 
biswarup@mri.ernet.in}, and
Sourov Roy$^{**}$\footnote{Electronic address: sourov@mri.ernet.in}}\\[5mm]
$^*$ Physics Department, Jadavpur University, Calcutta - 700 032, India\\
$^{**}$ Mehta Research Institute, Chhatnag Road, 
Jhusi, Allahabad - 221 506, India
\end{center}

\begin{abstract}

We show that the process $e^+e^-\lr \gamma+$ missing energy, arising from
the pair production of neutralinos, can probe the $\gamma-\tilde\gamma-
\grav$ as well as the $Z-\tilde Z-\grav$ couplings in Gauge Mediated 
Supersymmetry Breaking models. This enables one to study the mutual
relationship of the Goldstino couplings of the different gauginos, a feature
whose testability has not been emphasized so far. The Standard Model 
backgrounds get suppresed with the use of a right polarized electron beam.
The energy and angular distribution of the emitted photon can distinguish 
such models from the minimal supersymmetric theory and its variants.

\end{abstract}

\normalbaselineskip=28 true pt
\normalbaselines
\newpage
\setcounter{footnote}{0}
Distinguishing between various types of supersymmetry (SUSY) breaking schemes
is one crucial task ahead of those who are engaged in the search for SUSY in
nature \cite{susy}. 
Models with Gauge Mediated Supersymmetry Breaking (GMSB) \cite{gmsb1}
have recently attracted considerable attention in this 
context \cite{gmsb2,gmsb3,gmsb4,gmsb5}. The salient feature of such
models is that SUSY breaking is communicated radiatively  to the observable 
sector via ordinary gauge interactions, as opposed 
to more conventional theories where gravitation plays a decisive role. This is
usually implemented by introducing a messenger sector, consisting of 
vectorlike quark and lepton superfields $\hat\psi$.
The messenger superpotential contains a term of the form of
$\lambda S \hat{\bar\psi}\hat\psi$, where $S$ is a gauge singlet
superfield. The scalar and auxiliary components of $S$ acquire vacuum
expectation values (VEV) via interactions with the hidden sector.
This lifts the mass degeneracy between the messenger fermions and
sfermions. The observable gauginos
and sfermions acquire their masses at the one- and two-loop levels respectively,
with the messenger fields participating in the loops. The gaugino, squark and 
slepton masses are thus all related. An advantage of this scheme is that 
flavour-diagonal sfermion masses are induced at a rather low energy scale 
($\sim 100 $ TeV or so). This suppresses flavour-changing neutral currents
which could be potentially dangerous in supergravity type models due to 
interactions of heavy fields above the Grand Unification scale \cite{hall}.
Here the gravitino ($\grav$) becomes very light ($\sim 1$ eV). Therefore, the
lightest Standard Model superparticle can decay into it. Thus the so-called 
lightest SUSY particle (LSP) of the conventional model now becomes 
the next lightest
SUSY particle (NLSP), and is unstable; in fact, in the minimal version
of GMSB, it can decay within the detector
over a large part of the parameter space\footnote{As can be seen, for example,
from Bagger {\it et al.} in ref \cite{gmsb4}, a neutralino NLSP $\chi$ can 
decay within the detector so long as the parameter $M/\Lambda$ 
(defined later in the text) is
not appreciably higher than 100, and if $|\vec p_{\chi}|\sim m_{\chi}$.
The decay length can possibly be larger if the gravitino decay constant
gets contributions from additional fields present in the SUSY breaking sector,
as discussed by Ambrosanio {\it et al.} in ref. \cite{signal2}.},
making the observable signals radically
different from those of the minimal 
\newpage SUSY Standard Model (MSSM).

If there is one generation of messengers, the lightest neutralino is the NLSP. 
With a larger number of messenger generations, a right slepton may take that
place. Here we are considering the case with a neutralino NLSP, where the 
messenger fields form one complete $5+\bar 5$ of $SU(5)$. The entire mass
spectrum as well as the chargino and neutralino mixing matrices are determined
by the messenger mass scale $M$, and more crucially, by
a parameter $\L$ (the ratio
of the VEVs of the scalar and auxiliary components of $S$), together with the
Higgsino mass parameter $\mu$, $\tan\beta$ (the ratio of the two Higgs' VEVs)
and the number of messenger generations.

If a neutralino is the NLSP, it is almost always 
dominated by the Bino component.
The main decay mode of such an NLSP is $\cone\lr\gamma\grav$. Based on this, 
signals of GMSB have been predicted for colliders like the Fermilab
Tevatron, LEP-2 
and the Next Linear Collider (NLC) \cite{signal1,signal2,signal3}. 
However, the possibility of testing  
the components of $\cone$ other than the photino has so far been only
briefly commented upon (for example, by Dimopoulos {\it et al.} in
reference \cite{gmsb4} and Ambrosanio {\it et al.} in reference \cite{signal1}).
The Zino component of $\cone$ will give rise to the decay $\cone\lr Z\grav$
if $\mc >m_Z$, although the branching ratio will be rather small. If 
the signature of such a decay is available, then one can address two important
issues related to the structure of GMSB. First, it will probe the composition 
of the NLSP. Secondly, it will provide a test for the 
coupling of the $Z-\tilde Z$ current with the longitudinal component of
$\grav$, whose strength is related to similar coupling of the  
$\gamma-\tilde\gamma$ current \cite{fayet}.  In this paper we suggest
a new way of performing 
both these tests by considering the process
$e^+e^-\lr\cone\cone\lr\gamma\grav Z\grav$, followed by the $Z$ decaying 
invisibly into a pair of neutrinos. In this way, a small but finite Zino
component of the NLSP shows up in single photon events (SPE) in high-energy
$e^+e^-$ colliders, the NLC in particular. As we shall demonstrate here, such
events are easier to isolate from the 
backgrounds if polarized electron beams are used. 

The advantage of this channel is that SPEs are in any case observed with great 
care in $e^+e^-$ colliders for testing QED as well as physics beyond the 
Standard Models with heavy stable neutral particles. In the context of GMSB,
the signal proposed by us will also serve as 
a confirmatory test while more copious
two-photon events are likely to be the primary signal.

The production cross-section of $\cone$ pairs at high energy $e^+e^-$ colliders
as a function of the centre-of-mass energy $\sqrt{s}$ is presented in Figure 1
for both polarized and unpolarized electron beams. Here the controlling factors 
are the $\cone$ mass ($m_{\cone}$) and the neutralino mixing angles. These
are essentially determined by $\L$ throughout the parameter space of our 
interest. This is because in this region $\cone$ is dominantly Bino and 
hence the influence of $\mu$ or $\tan\beta$ is small. The only exception 
occurs if $|\mu|$ is small compared to the gaugino mass parameters, in which
case $\cone$ has a substantial Higgsino part. Then the $t$-channel 
slepton exchange cross section is suppressed 
by the electron mass and the signal 
becomes very weak. In all the numerical results presented here, we have used
$\mu=-300$ GeV, $\tan\beta=10$ and $M/\L=2$.

As long as the gravitino mass is negligible compared to the relevant energy
scale, it can be approximated by its Goldstino components.
Since the Goldstino coupling to a particle-sparticle pair is
universal in form, $Z-\tilde Z-\grav$ and $\gamma-\tilde\gamma-\grav$ couplings
are related \cite{gher}. 
$B(\cone\lr\gamma\grav)$ and $B(\cone\lr Z\grav)$ are thus computed
in a straightforward way. Some representative branching ratios are listed 
in Table 1. It can be seen from the table that over a large region of the 
parameter space $B(\cone\lr Z\grav)$ is small but non-negligible. For 
relatively large $m_{\cone}$, it can be as large as 23\%. However, in these 
cases the production cross-section is suppressed due to a correspondingly
large selectron mass which occurs in the $t$ and $u$-channel propagators. 

>From Figure 1 and Table 1, it is clear that at $\sqrt{s}=500$ GeV, the signal 
cross section without any kinematical cut is of the order of a few fb.
The main irreducible Standard Model background comes from $e^+e^-\lr\nu\bar\nu
\gamma$ whose cross section is 5.7 pb for unpolarized beams 
\cite{smback}. This is
large enough to swamp the signals.

The situation improves somewhat with the introduction of kinematical cuts.
Following \cite{ad} we have introduced a set of cuts as given below:
\be
95~{\rm GeV}<E_{\gamma}<225~{\rm GeV}, \ \ 40^o<\theta_\gamma<140^o.
\ee
These cuts also take care of the background from the radiative Bhabha 
scattering where the charged particles in the final states are lost in the
beam pipe \cite{chen}. 
The signal cross section with these cuts are plotted in Figure 2.
Only for a limited region of the parameter space the statistical significance
$S/\sqrt{B}$ rises above 2, where $S$ and $B$ are respectively the number
of signal and background events corresponding to an integrated luminosity
of 30 fb$^{-1}$. By strengthening the lower cut on $E_\gamma$ the significance
increases somewhat but is always well below 5. The situation does not improve
by using alternative sets of cuts, {\em e.g.}, the one proposed in \cite{chen}.
One, therefore, is forced to the conclusion that the situation is not 
optimistic for unpolarized beams. 

The signal improves drastically when polarized electron beams are available
\cite{pol}.
In this case, the irreducible Standard Model background from
$\nu \bar{\nu} \gamma$ is heavily suppressed
since the $t$-channel $W$-exchange 
diagrams do not contribute anymore. The signal,
on the other hand, is jacked up 
due to the larger hypercharge of the right-handed
gauge singlet electron coupling with the Bino-dominated NLSP. Using the above
cuts, some representative signal cross sections with 100\% right-polarized
electrons at $\sqrt{s}=500$ GeV are plotted in Figure 3. Also plotted are
the values corresponding to different significance levels. Here we have 
conservatively assumed an integrated luminosity of 10 fb$^{-1}$ with 
polarized beams. It is seen that for $130~{\rm GeV}\leq m_{\cone}\leq 210
~{\rm GeV}$, the signal is above the 5$\sigma$ level. We have also checked that
the other source of standard model backgrounds, namely, 
$e^{+}e^{-} \longrightarrow Z Z \gamma$ followed by both the Z's decaying
invisibly, is not significant for the polarized signals even at a $10 \sigma$
level \cite{barger}.  
Figures 2 and 3 also confirm the fact that the dependence of the signal
on $\mu$ is negligible for $|\mu|$ larger than the gaugino mass parameters.
It should be emphasized 
that an even larger region of the parameter space can be explored if (a) 
a larger $\sqrt{s}$ is available, and (b) one has a larger integrated 
luminosity. Thus SPEs provide us with a strong handle on the $Z-\tilde Z-
\grav$ interaction in the GMSB scenario.

However, the task of distinguishing GMSB from conventional SUSY scenarios
is still to be accomplished. This is important because the latter may also 
lead to an excess of SPEs at the NLC. The relevant processes there are
\cite{ad,pandita}
\bea
i)\  e^+e^-&\lr& \cone\cone\gamma\nonumber\\
ii)\ e^+e^-&\lr& \tilde\nu\tilde{\bar\nu}\gamma\nonumber\\
iii)\ e^+e^-&\lr& \chi^0_i\chi^0_j\gamma \ \ (i,j=1,2;\ i=j\not=1)\nonumber
\eea
Processes (ii) and (iii) occur in the special case where  $\tilde\nu$ 
and $\chi^0_2$ decay invisibly (the ``Virtual LSP" scenario \cite{vlsp}).

In Figure 4 we present the uncut cross sections with right polarized electron
beams for the above processes. The relevant formulae can be obtained 
from the Appendix of \cite{ad} by taking limits 
appropriate for right polarized beams. We have chosen the 
SUSY parameters in such a way
as to keep the MSSM contribution on the higher side so that the challenge
of distinguishing GMSB signals from them becomes most obvious. These 
include three cases where the invisible final state comes from the lightest
neutralino (LSP), the LSP and the $\tilde\nu$ and the LSP, the $\tilde
\nu$ and the second lightest neutralino.  
The $\tilde\nu$ contributes when its dominant decay is invisible through 
the channel $\tilde\nu\lr\cone\nu$. In such a case, $\chi^0_2$ may also 
become invisible if its dominant decay mode is 
$\chi^0_2\lr\nu\tilde{\bar\nu}$
It is seen from Figure 4 that the size of this signal is larger than that
of GMSB, and can very well mimic it. Thus, the size of the signal alone 
cannot be the deciding factor in discriminating between the two scenarios.

The two scenarios, however, {\em can} be distinguished from the energy
and angular distributions of the final state photon. We present in Figure 5
the $E_{\gamma}$ distributions in GMSB as well as in MSSM, subject to the
angular cut $40^o<\theta_\gamma<140^o$. The curve corresponding to GMSB
is almost flat and shows the end-points characteristic of two-body decays.
Moreover, it extends to higher values of $E_\gamma$ compared to the MSSM
distribution. On the other hand, the MSSM curves with different choices
of SUSY parameters show a continuous fall with energy, which can be clearly
distinguished from the previous case. 

In Figure 6, we show the angular distribution in the two scenarios 
subject to the energy cuts $95~{\rm GeV}<E_\gamma<225~{\rm GeV}$. The
MSSM distributions are well-populated for photons relatively closer
to the beam pipe with a distinct dip in the central region. On the 
contrary, the GMSB signal shows exactly the opposite trend. It may be
argued that the $Z-\tilde Z-\grav$ may also be probed by processes
where the $Z$ decays into a pair of jets. In such cases, the signal size
will definitely be larger. The dominant background in this case 
will come from $e^+e^-\lr q\bar q\gamma$. It may be naively argued that
a missing energy cut can discriminate the GMSB signal from the background.
However, jet fragmentation, mismeasurement of jet momenta, soft gluon
radiation etc. make the computation of the background much more 
cumbersome and uncertain. 

We have repeated the calculation by varying the degree of polarization
of the $e^-$ beam. Our main conclusions remain qualitatively unaffected
as long as more than 90\% right-polarized electron beams are available.

In conclusion, we reiterate that an excess of single photon events at
the NLC with polarized electron beams may provide a strong evidence 
for the $Z-\tilde Z-\grav$ coupling which is an inescapable consequence
of the GMSB scenario. The Zino component of the NLSP can also be tested
by the same signal. The signal is distinguishable not only from the
Standard Model background but also from similar signals arising in
MSSM without a light gravitino.

\vskip 0.5 true cm

Amitava Datta's work is supported by  
the Department of Science and
Technology, Government of India, under grant no. SP/S2/K-07/92. 
The work of Aseshkrishna Datta has been supported by the 
Council for Scientific and Industrial Research, India.
AK and SR acknowledge the hospitality of the Theory Group, Saha Institute
of Nuclear Physics, Calcutta, and BM and SR thank the Physics Department,
Jadavpur University, where part of this work was done.

\newpage

\begin{center}
\begin{tabular}{|c|c|c|c|} 
\hline
& & & \cr
M (TeV)& $\mc$ (GeV)& $B(\cone\r\gamma\grav)$&
$B(\cone\r Z\grav)$\cr
& & & \cr
\hline
& & & \cr
 160&  111.5&  0.996&  0.004\cr
 190&  132.0&  0.975&  0.025\cr
 220&  152.4&  0.943&  0.057\cr
 250&  172.5&  0.910&  0.090\cr
 280&  192.3&  0.878&  0.122\cr
 310&  211.5&  0.846&  0.154\cr
 340&  229.8&  0.812&  0.188\cr
 370&  246.4&  0.767&  0.233\cr
& & & \cr
\hline
\end{tabular}
\end{center}

\vskip 1 true cm
\centerline{\bf Table 1}

\vskip 2 true cm

\noindent The branching ratios of the lightest neutralino that can be 
produced in the NLC with $\sqrt{s}=500$ GeV, to a photon
and to a $Z$ (alongwith a $\grav$ in each case). We have chosen 
$\mu=-300$ GeV, $\tan\beta=10$ and $M/\L =2$. The decay of the NLSP
to a Higgs is negligible.   

\newpage
\normalbaselineskip = 16 true pt
\normalbaselines
\setlength{\unitlength}{0.240900pt}
\ifx\plotpoint\undefined\newsavebox{\plotpoint}\fi
\sbox{\plotpoint}{\rule[-0.200pt]{0.400pt}{0.400pt}}%
\begin{picture}(1500,900)(0,0)
\font\gnuplot=cmr10 at 10pt
\gnuplot
\sbox{\plotpoint}{\rule[-0.200pt]{0.400pt}{0.400pt}}%
\put(220.0,113.0){\rule[-0.200pt]{4.818pt}{0.400pt}}
\put(198,113){\makebox(0,0)[r]{0.0001}}
\put(1416.0,113.0){\rule[-0.200pt]{4.818pt}{0.400pt}}
\put(220.0,159.0){\rule[-0.200pt]{2.409pt}{0.400pt}}
\put(1426.0,159.0){\rule[-0.200pt]{2.409pt}{0.400pt}}
\put(220.0,220.0){\rule[-0.200pt]{2.409pt}{0.400pt}}
\put(1426.0,220.0){\rule[-0.200pt]{2.409pt}{0.400pt}}
\put(220.0,251.0){\rule[-0.200pt]{2.409pt}{0.400pt}}
\put(1426.0,251.0){\rule[-0.200pt]{2.409pt}{0.400pt}}
\put(220.0,266.0){\rule[-0.200pt]{4.818pt}{0.400pt}}
\put(198,266){\makebox(0,0)[r]{0.001}}
\put(1416.0,266.0){\rule[-0.200pt]{4.818pt}{0.400pt}}
\put(220.0,312.0){\rule[-0.200pt]{2.409pt}{0.400pt}}
\put(1426.0,312.0){\rule[-0.200pt]{2.409pt}{0.400pt}}
\put(220.0,373.0){\rule[-0.200pt]{2.409pt}{0.400pt}}
\put(1426.0,373.0){\rule[-0.200pt]{2.409pt}{0.400pt}}
\put(220.0,404.0){\rule[-0.200pt]{2.409pt}{0.400pt}}
\put(1426.0,404.0){\rule[-0.200pt]{2.409pt}{0.400pt}}
\put(220.0,419.0){\rule[-0.200pt]{4.818pt}{0.400pt}}
\put(198,419){\makebox(0,0)[r]{0.01}}
\put(1416.0,419.0){\rule[-0.200pt]{4.818pt}{0.400pt}}
\put(220.0,465.0){\rule[-0.200pt]{2.409pt}{0.400pt}}
\put(1426.0,465.0){\rule[-0.200pt]{2.409pt}{0.400pt}}
\put(220.0,525.0){\rule[-0.200pt]{2.409pt}{0.400pt}}
\put(1426.0,525.0){\rule[-0.200pt]{2.409pt}{0.400pt}}
\put(220.0,557.0){\rule[-0.200pt]{2.409pt}{0.400pt}}
\put(1426.0,557.0){\rule[-0.200pt]{2.409pt}{0.400pt}}
\put(220.0,571.0){\rule[-0.200pt]{4.818pt}{0.400pt}}
\put(198,571){\makebox(0,0)[r]{0.1}}
\put(1416.0,571.0){\rule[-0.200pt]{4.818pt}{0.400pt}}
\put(220.0,617.0){\rule[-0.200pt]{2.409pt}{0.400pt}}
\put(1426.0,617.0){\rule[-0.200pt]{2.409pt}{0.400pt}}
\put(220.0,678.0){\rule[-0.200pt]{2.409pt}{0.400pt}}
\put(1426.0,678.0){\rule[-0.200pt]{2.409pt}{0.400pt}}
\put(220.0,709.0){\rule[-0.200pt]{2.409pt}{0.400pt}}
\put(1426.0,709.0){\rule[-0.200pt]{2.409pt}{0.400pt}}
\put(220.0,724.0){\rule[-0.200pt]{4.818pt}{0.400pt}}
\put(198,724){\makebox(0,0)[r]{1}}
\put(1416.0,724.0){\rule[-0.200pt]{4.818pt}{0.400pt}}
\put(220.0,770.0){\rule[-0.200pt]{2.409pt}{0.400pt}}
\put(1426.0,770.0){\rule[-0.200pt]{2.409pt}{0.400pt}}
\put(220.0,831.0){\rule[-0.200pt]{2.409pt}{0.400pt}}
\put(1426.0,831.0){\rule[-0.200pt]{2.409pt}{0.400pt}}
\put(220.0,862.0){\rule[-0.200pt]{2.409pt}{0.400pt}}
\put(1426.0,862.0){\rule[-0.200pt]{2.409pt}{0.400pt}}
\put(220.0,877.0){\rule[-0.200pt]{4.818pt}{0.400pt}}
\put(198,877){\makebox(0,0)[r]{10}}
\put(1416.0,877.0){\rule[-0.200pt]{4.818pt}{0.400pt}}
\put(220.0,113.0){\rule[-0.200pt]{0.400pt}{4.818pt}}
\put(220,68){\makebox(0,0){300}}
\put(220.0,857.0){\rule[-0.200pt]{0.400pt}{4.818pt}}
\put(394.0,113.0){\rule[-0.200pt]{0.400pt}{4.818pt}}
\put(394,68){\makebox(0,0){400}}
\put(394.0,857.0){\rule[-0.200pt]{0.400pt}{4.818pt}}
\put(567.0,113.0){\rule[-0.200pt]{0.400pt}{4.818pt}}
\put(567,68){\makebox(0,0){500}}
\put(567.0,857.0){\rule[-0.200pt]{0.400pt}{4.818pt}}
\put(741.0,113.0){\rule[-0.200pt]{0.400pt}{4.818pt}}
\put(741,68){\makebox(0,0){600}}
\put(741.0,857.0){\rule[-0.200pt]{0.400pt}{4.818pt}}
\put(915.0,113.0){\rule[-0.200pt]{0.400pt}{4.818pt}}
\put(915,68){\makebox(0,0){700}}
\put(915.0,857.0){\rule[-0.200pt]{0.400pt}{4.818pt}}
\put(1089.0,113.0){\rule[-0.200pt]{0.400pt}{4.818pt}}
\put(1089,68){\makebox(0,0){800}}
\put(1089.0,857.0){\rule[-0.200pt]{0.400pt}{4.818pt}}
\put(1262.0,113.0){\rule[-0.200pt]{0.400pt}{4.818pt}}
\put(1262,68){\makebox(0,0){900}}
\put(1262.0,857.0){\rule[-0.200pt]{0.400pt}{4.818pt}}
\put(1436.0,113.0){\rule[-0.200pt]{0.400pt}{4.818pt}}
\put(1436,68){\makebox(0,0){1000}}
\put(1436.0,857.0){\rule[-0.200pt]{0.400pt}{4.818pt}}
\put(220.0,113.0){\rule[-0.200pt]{292.934pt}{0.400pt}}
\put(1436.0,113.0){\rule[-0.200pt]{0.400pt}{184.048pt}}
\put(220.0,877.0){\rule[-0.200pt]{292.934pt}{0.400pt}}
\put(45,495){\makebox(0,0){$\sigma$ (pb)}}
\put(828,23){\makebox(0,0){$E_{cm}$ (GeV)}}
\put(220.0,113.0){\rule[-0.200pt]{0.400pt}{184.048pt}}
\put(1306,812){\makebox(0,0)[r]{"R-polarized $e^-$"}}
\put(1328.0,812.0){\rule[-0.200pt]{15.899pt}{0.400pt}}
\multiput(299.58,572.00)(0.498,0.552){93}{\rule{0.120pt}{0.542pt}}
\multiput(298.17,572.00)(48.000,51.876){2}{\rule{0.400pt}{0.271pt}}
\multiput(347.00,625.58)(0.945,0.497){47}{\rule{0.852pt}{0.120pt}}
\multiput(347.00,624.17)(45.232,25.000){2}{\rule{0.426pt}{0.400pt}}
\multiput(394.00,650.58)(1.709,0.494){25}{\rule{1.443pt}{0.119pt}}
\multiput(394.00,649.17)(44.005,14.000){2}{\rule{0.721pt}{0.400pt}}
\multiput(441.00,664.59)(3.135,0.488){13}{\rule{2.500pt}{0.117pt}}
\multiput(441.00,663.17)(42.811,8.000){2}{\rule{1.250pt}{0.400pt}}
\multiput(489.00,672.59)(5.163,0.477){7}{\rule{3.860pt}{0.115pt}}
\multiput(489.00,671.17)(38.988,5.000){2}{\rule{1.930pt}{0.400pt}}
\multiput(536.00,677.61)(10.286,0.447){3}{\rule{6.367pt}{0.108pt}}
\multiput(536.00,676.17)(33.786,3.000){2}{\rule{3.183pt}{0.400pt}}
\put(583,679.67){\rule{11.563pt}{0.400pt}}
\multiput(583.00,679.17)(24.000,1.000){2}{\rule{5.782pt}{0.400pt}}
\put(631,680.67){\rule{11.322pt}{0.400pt}}
\multiput(631.00,680.17)(23.500,1.000){2}{\rule{5.661pt}{0.400pt}}
\put(678,680.67){\rule{11.563pt}{0.400pt}}
\multiput(678.00,681.17)(24.000,-1.000){2}{\rule{5.782pt}{0.400pt}}
\put(726,679.67){\rule{11.322pt}{0.400pt}}
\multiput(726.00,680.17)(23.500,-1.000){2}{\rule{5.661pt}{0.400pt}}
\put(773,678.67){\rule{11.322pt}{0.400pt}}
\multiput(773.00,679.17)(23.500,-1.000){2}{\rule{5.661pt}{0.400pt}}
\put(820,677.17){\rule{9.700pt}{0.400pt}}
\multiput(820.00,678.17)(27.867,-2.000){2}{\rule{4.850pt}{0.400pt}}
\put(868,675.17){\rule{9.500pt}{0.400pt}}
\multiput(868.00,676.17)(27.282,-2.000){2}{\rule{4.750pt}{0.400pt}}
\put(915,673.17){\rule{9.500pt}{0.400pt}}
\multiput(915.00,674.17)(27.282,-2.000){2}{\rule{4.750pt}{0.400pt}}
\put(962,671.17){\rule{9.700pt}{0.400pt}}
\multiput(962.00,672.17)(27.867,-2.000){2}{\rule{4.850pt}{0.400pt}}
\put(1010,669.17){\rule{9.500pt}{0.400pt}}
\multiput(1010.00,670.17)(27.282,-2.000){2}{\rule{4.750pt}{0.400pt}}
\multiput(1057.00,667.95)(10.286,-0.447){3}{\rule{6.367pt}{0.108pt}}
\multiput(1057.00,668.17)(33.786,-3.000){2}{\rule{3.183pt}{0.400pt}}
\put(1104,664.17){\rule{9.700pt}{0.400pt}}
\multiput(1104.00,665.17)(27.867,-2.000){2}{\rule{4.850pt}{0.400pt}}
\multiput(1152.00,662.95)(10.286,-0.447){3}{\rule{6.367pt}{0.108pt}}
\multiput(1152.00,663.17)(33.786,-3.000){2}{\rule{3.183pt}{0.400pt}}
\put(1199,659.17){\rule{9.700pt}{0.400pt}}
\multiput(1199.00,660.17)(27.867,-2.000){2}{\rule{4.850pt}{0.400pt}}
\multiput(1247.00,657.95)(10.286,-0.447){3}{\rule{6.367pt}{0.108pt}}
\multiput(1247.00,658.17)(33.786,-3.000){2}{\rule{3.183pt}{0.400pt}}
\put(1294,654.17){\rule{9.500pt}{0.400pt}}
\multiput(1294.00,655.17)(27.282,-2.000){2}{\rule{4.750pt}{0.400pt}}
\multiput(1341.00,652.95)(10.509,-0.447){3}{\rule{6.500pt}{0.108pt}}
\multiput(1341.00,653.17)(34.509,-3.000){2}{\rule{3.250pt}{0.400pt}}
\put(1389,649.17){\rule{9.500pt}{0.400pt}}
\multiput(1389.00,650.17)(27.282,-2.000){2}{\rule{4.750pt}{0.400pt}}
\put(299.0,113.0){\rule[-0.200pt]{0.400pt}{110.573pt}}
\sbox{\plotpoint}{\rule[-0.500pt]{1.000pt}{1.000pt}}%
\put(1306,767){\makebox(0,0)[r]{"L-polarized $e^-$"}}
\multiput(1328,767)(20.756,0.000){4}{\usebox{\plotpoint}}
\put(1394,767){\usebox{\plotpoint}}
\multiput(299,113)(0.000,20.756){8}{\usebox{\plotpoint}}
\multiput(299,262)(12.966,16.207){3}{\usebox{\plotpoint}}
\multiput(347,322)(17.156,11.681){3}{\usebox{\plotpoint}}
\multiput(394,354)(19.098,8.127){3}{\usebox{\plotpoint}}
\multiput(441,374)(19.925,5.812){2}{\usebox{\plotpoint}}
\multiput(489,388)(20.301,4.319){2}{\usebox{\plotpoint}}
\multiput(536,398)(20.461,3.483){3}{\usebox{\plotpoint}}
\multiput(583,406)(20.644,2.150){2}{\usebox{\plotpoint}}
\multiput(631,411)(20.681,1.760){2}{\usebox{\plotpoint}}
\multiput(678,415)(20.715,1.295){3}{\usebox{\plotpoint}}
\multiput(726,418)(20.713,1.322){2}{\usebox{\plotpoint}}
\multiput(773,421)(20.751,0.442){2}{\usebox{\plotpoint}}
\multiput(820,422)(20.751,0.432){2}{\usebox{\plotpoint}}
\multiput(868,423)(20.751,0.442){3}{\usebox{\plotpoint}}
\multiput(915,424)(20.756,0.000){2}{\usebox{\plotpoint}}
\multiput(962,424)(20.756,0.000){2}{\usebox{\plotpoint}}
\multiput(1010,424)(20.756,0.000){2}{\usebox{\plotpoint}}
\multiput(1057,424)(20.751,-0.442){3}{\usebox{\plotpoint}}
\multiput(1104,423)(20.751,-0.432){2}{\usebox{\plotpoint}}
\multiput(1152,422)(20.756,0.000){2}{\usebox{\plotpoint}}
\multiput(1199,422)(20.751,-0.432){3}{\usebox{\plotpoint}}
\multiput(1247,421)(20.737,-0.882){2}{\usebox{\plotpoint}}
\multiput(1294,419)(20.751,-0.442){2}{\usebox{\plotpoint}}
\multiput(1341,418)(20.751,-0.432){2}{\usebox{\plotpoint}}
\multiput(1389,417)(20.751,-0.442){3}{\usebox{\plotpoint}}
\put(1436,416){\usebox{\plotpoint}}
\sbox{\plotpoint}{\rule[-0.400pt]{0.800pt}{0.800pt}}%
\put(1306,722){\makebox(0,0)[r]{"Unpolarized $e^-$"}}
\put(1328.0,722.0){\rule[-0.400pt]{15.899pt}{0.800pt}}
\multiput(300.41,527.00)(0.502,0.551){89}{\rule{0.121pt}{1.083pt}}
\multiput(297.34,527.00)(48.000,50.751){2}{\rule{0.800pt}{0.542pt}}
\multiput(347.00,581.41)(0.992,0.504){41}{\rule{1.767pt}{0.122pt}}
\multiput(347.00,578.34)(43.333,24.000){2}{\rule{0.883pt}{0.800pt}}
\multiput(394.00,605.41)(1.626,0.508){23}{\rule{2.707pt}{0.122pt}}
\multiput(394.00,602.34)(41.382,15.000){2}{\rule{1.353pt}{0.800pt}}
\multiput(441.00,620.40)(3.406,0.520){9}{\rule{5.000pt}{0.125pt}}
\multiput(441.00,617.34)(37.622,8.000){2}{\rule{2.500pt}{0.800pt}}
\multiput(489.00,628.38)(7.477,0.560){3}{\rule{7.720pt}{0.135pt}}
\multiput(489.00,625.34)(30.977,5.000){2}{\rule{3.860pt}{0.800pt}}
\put(536,631.84){\rule{11.322pt}{0.800pt}}
\multiput(536.00,630.34)(23.500,3.000){2}{\rule{5.661pt}{0.800pt}}
\put(583,633.84){\rule{11.563pt}{0.800pt}}
\multiput(583.00,633.34)(24.000,1.000){2}{\rule{5.782pt}{0.800pt}}
\put(631,634.84){\rule{11.322pt}{0.800pt}}
\multiput(631.00,634.34)(23.500,1.000){2}{\rule{5.661pt}{0.800pt}}
\put(678,634.84){\rule{11.563pt}{0.800pt}}
\multiput(678.00,635.34)(24.000,-1.000){2}{\rule{5.782pt}{0.800pt}}
\put(299.0,113.0){\rule[-0.400pt]{0.800pt}{99.733pt}}
\put(773,633.34){\rule{11.322pt}{0.800pt}}
\multiput(773.00,634.34)(23.500,-2.000){2}{\rule{5.661pt}{0.800pt}}
\put(820,631.84){\rule{11.563pt}{0.800pt}}
\multiput(820.00,632.34)(24.000,-1.000){2}{\rule{5.782pt}{0.800pt}}
\put(868,630.34){\rule{11.322pt}{0.800pt}}
\multiput(868.00,631.34)(23.500,-2.000){2}{\rule{5.661pt}{0.800pt}}
\put(915,628.34){\rule{11.322pt}{0.800pt}}
\multiput(915.00,629.34)(23.500,-2.000){2}{\rule{5.661pt}{0.800pt}}
\put(962,625.84){\rule{11.563pt}{0.800pt}}
\multiput(962.00,627.34)(24.000,-3.000){2}{\rule{5.782pt}{0.800pt}}
\put(1010,623.34){\rule{11.322pt}{0.800pt}}
\multiput(1010.00,624.34)(23.500,-2.000){2}{\rule{5.661pt}{0.800pt}}
\put(1057,621.34){\rule{11.322pt}{0.800pt}}
\multiput(1057.00,622.34)(23.500,-2.000){2}{\rule{5.661pt}{0.800pt}}
\put(1104,618.84){\rule{11.563pt}{0.800pt}}
\multiput(1104.00,620.34)(24.000,-3.000){2}{\rule{5.782pt}{0.800pt}}
\put(1152,616.34){\rule{11.322pt}{0.800pt}}
\multiput(1152.00,617.34)(23.500,-2.000){2}{\rule{5.661pt}{0.800pt}}
\put(1199,613.84){\rule{11.563pt}{0.800pt}}
\multiput(1199.00,615.34)(24.000,-3.000){2}{\rule{5.782pt}{0.800pt}}
\put(1247,611.34){\rule{11.322pt}{0.800pt}}
\multiput(1247.00,612.34)(23.500,-2.000){2}{\rule{5.661pt}{0.800pt}}
\put(1294,609.34){\rule{11.322pt}{0.800pt}}
\multiput(1294.00,610.34)(23.500,-2.000){2}{\rule{5.661pt}{0.800pt}}
\put(1341,606.84){\rule{11.563pt}{0.800pt}}
\multiput(1341.00,608.34)(24.000,-3.000){2}{\rule{5.782pt}{0.800pt}}
\put(1389,604.34){\rule{11.322pt}{0.800pt}}
\multiput(1389.00,605.34)(23.500,-2.000){2}{\rule{5.661pt}{0.800pt}}
\put(726.0,636.0){\rule[-0.400pt]{11.322pt}{0.800pt}}
\end{picture}
\vskip 2 true cm
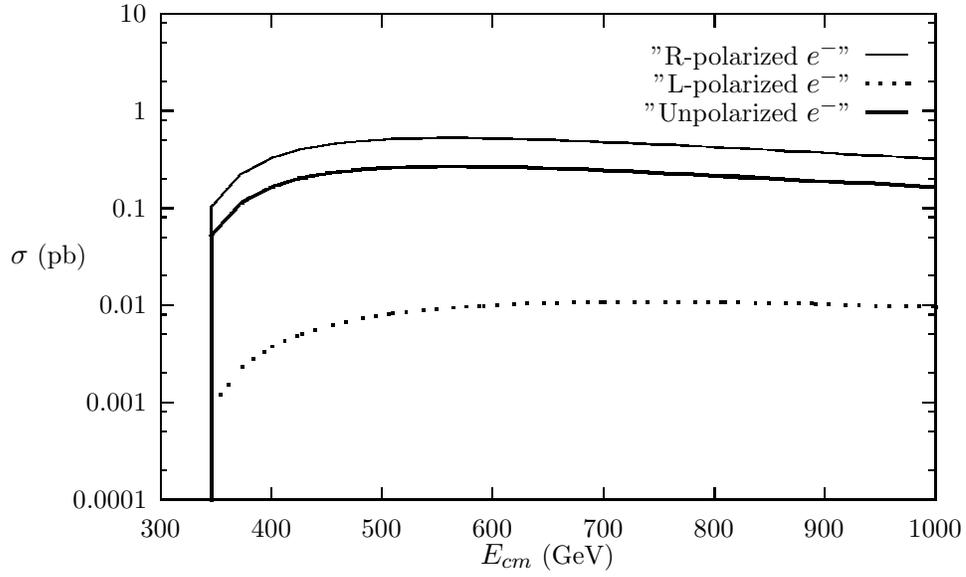
\centerline{\bf Fig. 1}

\noindent The cross-section for $e^+e^-\longrightarrow \chi^0_1\chi^0_1$
against the center-of-mass energy, for $\mu=-300$ GeV, $\tan\beta=10$,
$M=230000$ GeV and $M/\Lambda=2$ (which yields $m_{\chi_1^0}=159$ GeV).
\newpage

%
%
\setlength{\unitlength}{0.240900pt}
\ifx\plotpoint\undefined\newsavebox{\plotpoint}\fi
\sbox{\plotpoint}{\rule[-0.200pt]{0.400pt}{0.400pt}}%
\begin{picture}(1500,900)(0,0)
\font\gnuplot=cmr10 at 10pt
\gnuplot
\sbox{\plotpoint}{\rule[-0.200pt]{0.400pt}{0.400pt}}%
\put(220.0,113.0){\rule[-0.200pt]{292.934pt}{0.400pt}}
\put(220.0,113.0){\rule[-0.200pt]{4.818pt}{0.400pt}}
\put(198,113){\makebox(0,0)[r]{0}}
\put(1416.0,113.0){\rule[-0.200pt]{4.818pt}{0.400pt}}
\put(220.0,209.0){\rule[-0.200pt]{4.818pt}{0.400pt}}
\put(198,209){\makebox(0,0)[r]{0.001}}
\put(1416.0,209.0){\rule[-0.200pt]{4.818pt}{0.400pt}}
\put(220.0,304.0){\rule[-0.200pt]{4.818pt}{0.400pt}}
\put(198,304){\makebox(0,0)[r]{0.002}}
\put(1416.0,304.0){\rule[-0.200pt]{4.818pt}{0.400pt}}
\put(220.0,400.0){\rule[-0.200pt]{4.818pt}{0.400pt}}
\put(198,400){\makebox(0,0)[r]{0.003}}
\put(1416.0,400.0){\rule[-0.200pt]{4.818pt}{0.400pt}}
\put(220.0,495.0){\rule[-0.200pt]{4.818pt}{0.400pt}}
\put(198,495){\makebox(0,0)[r]{0.004}}
\put(1416.0,495.0){\rule[-0.200pt]{4.818pt}{0.400pt}}
\put(220.0,591.0){\rule[-0.200pt]{4.818pt}{0.400pt}}
\put(198,591){\makebox(0,0)[r]{0.005}}
\put(1416.0,591.0){\rule[-0.200pt]{4.818pt}{0.400pt}}
\put(220.0,686.0){\rule[-0.200pt]{4.818pt}{0.400pt}}
\put(198,686){\makebox(0,0)[r]{0.006}}
\put(1416.0,686.0){\rule[-0.200pt]{4.818pt}{0.400pt}}
\put(220.0,782.0){\rule[-0.200pt]{4.818pt}{0.400pt}}
\put(198,782){\makebox(0,0)[r]{0.007}}
\put(1416.0,782.0){\rule[-0.200pt]{4.818pt}{0.400pt}}
\put(220.0,877.0){\rule[-0.200pt]{4.818pt}{0.400pt}}
\put(198,877){\makebox(0,0)[r]{0.008}}
\put(1416.0,877.0){\rule[-0.200pt]{4.818pt}{0.400pt}}
\put(220.0,113.0){\rule[-0.200pt]{0.400pt}{4.818pt}}
\put(220,68){\makebox(0,0){100}}
\put(220.0,857.0){\rule[-0.200pt]{0.400pt}{4.818pt}}
\put(372.0,113.0){\rule[-0.200pt]{0.400pt}{4.818pt}}
\put(372,68){\makebox(0,0){120}}
\put(372.0,857.0){\rule[-0.200pt]{0.400pt}{4.818pt}}
\put(524.0,113.0){\rule[-0.200pt]{0.400pt}{4.818pt}}
\put(524,68){\makebox(0,0){140}}
\put(524.0,857.0){\rule[-0.200pt]{0.400pt}{4.818pt}}
\put(676.0,113.0){\rule[-0.200pt]{0.400pt}{4.818pt}}
\put(676,68){\makebox(0,0){160}}
\put(676.0,857.0){\rule[-0.200pt]{0.400pt}{4.818pt}}
\put(828.0,113.0){\rule[-0.200pt]{0.400pt}{4.818pt}}
\put(828,68){\makebox(0,0){180}}
\put(828.0,857.0){\rule[-0.200pt]{0.400pt}{4.818pt}}
\put(980.0,113.0){\rule[-0.200pt]{0.400pt}{4.818pt}}
\put(980,68){\makebox(0,0){200}}
\put(980.0,857.0){\rule[-0.200pt]{0.400pt}{4.818pt}}
\put(1132.0,113.0){\rule[-0.200pt]{0.400pt}{4.818pt}}
\put(1132,68){\makebox(0,0){220}}
\put(1132.0,857.0){\rule[-0.200pt]{0.400pt}{4.818pt}}
\put(1284.0,113.0){\rule[-0.200pt]{0.400pt}{4.818pt}}
\put(1284,68){\makebox(0,0){240}}
\put(1284.0,857.0){\rule[-0.200pt]{0.400pt}{4.818pt}}
\put(1436.0,113.0){\rule[-0.200pt]{0.400pt}{4.818pt}}
\put(1436,68){\makebox(0,0){260}}
\put(1436.0,857.0){\rule[-0.200pt]{0.400pt}{4.818pt}}
\put(220.0,113.0){\rule[-0.200pt]{292.934pt}{0.400pt}}
\put(1436.0,113.0){\rule[-0.200pt]{0.400pt}{184.048pt}}
\put(220.0,877.0){\rule[-0.200pt]{292.934pt}{0.400pt}}
\put(45,495){\makebox(0,0){$\sigma$ (pb)}}
\put(828,3){\makebox(0,0){$m_{NLSP}$ (GeV)}}
\put(220.0,113.0){\rule[-0.200pt]{0.400pt}{184.048pt}}
\put(1306,812){\makebox(0,0)[r]{"$\mu=-300$ GeV"}}
\put(1328.0,812.0){\rule[-0.200pt]{15.899pt}{0.400pt}}
\put(256,124){\usebox{\plotpoint}}
\multiput(256.00,124.58)(0.968,0.497){51}{\rule{0.870pt}{0.120pt}}
\multiput(256.00,123.17)(50.194,27.000){2}{\rule{0.435pt}{0.400pt}}
\multiput(308.00,151.58)(0.605,0.498){83}{\rule{0.584pt}{0.120pt}}
\multiput(308.00,150.17)(50.788,43.000){2}{\rule{0.292pt}{0.400pt}}
\multiput(360.58,194.00)(0.498,0.519){101}{\rule{0.120pt}{0.515pt}}
\multiput(359.17,194.00)(52.000,52.930){2}{\rule{0.400pt}{0.258pt}}
\multiput(412.58,248.00)(0.498,0.548){101}{\rule{0.120pt}{0.538pt}}
\multiput(411.17,248.00)(52.000,55.882){2}{\rule{0.400pt}{0.269pt}}
\multiput(464.58,305.00)(0.498,0.519){99}{\rule{0.120pt}{0.516pt}}
\multiput(463.17,305.00)(51.000,51.930){2}{\rule{0.400pt}{0.258pt}}
\multiput(515.00,358.58)(0.565,0.498){89}{\rule{0.552pt}{0.120pt}}
\multiput(515.00,357.17)(50.854,46.000){2}{\rule{0.276pt}{0.400pt}}
\multiput(567.00,404.58)(0.751,0.498){65}{\rule{0.700pt}{0.120pt}}
\multiput(567.00,403.17)(49.547,34.000){2}{\rule{0.350pt}{0.400pt}}
\multiput(618.00,438.58)(1.550,0.495){31}{\rule{1.324pt}{0.119pt}}
\multiput(618.00,437.17)(49.253,17.000){2}{\rule{0.662pt}{0.400pt}}
\multiput(670.00,455.59)(5.608,0.477){7}{\rule{4.180pt}{0.115pt}}
\multiput(670.00,454.17)(42.324,5.000){2}{\rule{2.090pt}{0.400pt}}
\multiput(721.00,458.93)(4.468,-0.482){9}{\rule{3.433pt}{0.116pt}}
\multiput(721.00,459.17)(42.874,-6.000){2}{\rule{1.717pt}{0.400pt}}
\multiput(771.00,452.92)(1.728,-0.494){27}{\rule{1.460pt}{0.119pt}}
\multiput(771.00,453.17)(47.970,-15.000){2}{\rule{0.730pt}{0.400pt}}
\multiput(822.00,437.92)(1.095,-0.496){43}{\rule{0.970pt}{0.120pt}}
\multiput(822.00,438.17)(47.988,-23.000){2}{\rule{0.485pt}{0.400pt}}
\multiput(872.00,414.92)(0.809,-0.497){59}{\rule{0.745pt}{0.120pt}}
\multiput(872.00,415.17)(48.453,-31.000){2}{\rule{0.373pt}{0.400pt}}
\multiput(922.00,383.92)(0.767,-0.497){61}{\rule{0.713pt}{0.120pt}}
\multiput(922.00,384.17)(47.521,-32.000){2}{\rule{0.356pt}{0.400pt}}
\multiput(971.00,351.92)(0.663,-0.498){71}{\rule{0.630pt}{0.120pt}}
\multiput(971.00,352.17)(47.693,-37.000){2}{\rule{0.315pt}{0.400pt}}
\multiput(1020.00,314.92)(0.649,-0.498){71}{\rule{0.619pt}{0.120pt}}
\multiput(1020.00,315.17)(46.715,-37.000){2}{\rule{0.309pt}{0.400pt}}
\multiput(1068.00,277.92)(0.653,-0.498){69}{\rule{0.622pt}{0.120pt}}
\multiput(1068.00,278.17)(45.709,-36.000){2}{\rule{0.311pt}{0.400pt}}
\multiput(1115.00,241.92)(0.692,-0.498){65}{\rule{0.653pt}{0.120pt}}
\multiput(1115.00,242.17)(45.645,-34.000){2}{\rule{0.326pt}{0.400pt}}
\multiput(1162.00,207.92)(0.752,-0.497){57}{\rule{0.700pt}{0.120pt}}
\multiput(1162.00,208.17)(43.547,-30.000){2}{\rule{0.350pt}{0.400pt}}
\multiput(1207.00,177.92)(0.884,-0.497){47}{\rule{0.804pt}{0.120pt}}
\multiput(1207.00,178.17)(42.331,-25.000){2}{\rule{0.402pt}{0.400pt}}
\multiput(1251.00,152.92)(1.007,-0.496){39}{\rule{0.900pt}{0.119pt}}
\multiput(1251.00,153.17)(40.132,-21.000){2}{\rule{0.450pt}{0.400pt}}
\multiput(1293.00,131.92)(1.352,-0.494){27}{\rule{1.167pt}{0.119pt}}
\multiput(1293.00,132.17)(37.579,-15.000){2}{\rule{0.583pt}{0.400pt}}
\put(220,409){\usebox{\plotpoint}}
\multiput(220,409)(20.756,0.000){59}{\usebox{\plotpoint}}
\put(1436,409){\usebox{\plotpoint}}
\put(400,440){$2\sigma$}
\sbox{\plotpoint}{\rule[-0.400pt]{0.800pt}{0.800pt}}%
\put(1306,722){\makebox(0,0)[r]{"$\mu=900$ GeV"}}
\put(1328.0,722.0){\rule[-0.400pt]{15.899pt}{0.800pt}}
\put(259,126){\usebox{\plotpoint}}
\multiput(259.00,127.41)(0.940,0.504){51}{\rule{1.690pt}{0.121pt}}
\multiput(259.00,124.34)(50.493,29.000){2}{\rule{0.845pt}{0.800pt}}
\multiput(313.00,156.41)(0.576,0.502){85}{\rule{1.122pt}{0.121pt}}
\multiput(313.00,153.34)(50.672,46.000){2}{\rule{0.561pt}{0.800pt}}
\multiput(367.41,201.00)(0.502,0.517){101}{\rule{0.121pt}{1.030pt}}
\multiput(364.34,201.00)(54.000,53.863){2}{\rule{0.800pt}{0.515pt}}
\multiput(421.41,257.00)(0.502,0.546){99}{\rule{0.121pt}{1.075pt}}
\multiput(418.34,257.00)(53.000,55.768){2}{\rule{0.800pt}{0.538pt}}
\multiput(473.00,316.41)(0.498,0.502){101}{\rule{1.000pt}{0.121pt}}
\multiput(473.00,313.34)(51.924,54.000){2}{\rule{0.500pt}{0.800pt}}
\multiput(527.00,370.41)(0.589,0.502){83}{\rule{1.142pt}{0.121pt}}
\multiput(527.00,367.34)(50.629,45.000){2}{\rule{0.571pt}{0.800pt}}
\multiput(580.00,415.41)(0.878,0.503){55}{\rule{1.594pt}{0.121pt}}
\multiput(580.00,412.34)(50.693,31.000){2}{\rule{0.797pt}{0.800pt}}
\multiput(634.00,446.41)(1.838,0.508){23}{\rule{3.027pt}{0.122pt}}
\multiput(634.00,443.34)(46.718,15.000){2}{\rule{1.513pt}{0.800pt}}
\put(687,459.34){\rule{13.009pt}{0.800pt}}
\multiput(687.00,458.34)(27.000,2.000){2}{\rule{6.504pt}{0.800pt}}
\multiput(741.00,460.08)(3.261,-0.516){11}{\rule{4.911pt}{0.124pt}}
\multiput(741.00,460.34)(42.807,-9.000){2}{\rule{2.456pt}{0.800pt}}
\multiput(794.00,451.09)(1.543,-0.506){29}{\rule{2.600pt}{0.122pt}}
\multiput(794.00,451.34)(48.604,-18.000){2}{\rule{1.300pt}{0.800pt}}
\multiput(848.00,433.09)(1.032,-0.504){45}{\rule{1.831pt}{0.121pt}}
\multiput(848.00,433.34)(49.200,-26.000){2}{\rule{0.915pt}{0.800pt}}
\multiput(901.00,407.09)(0.823,-0.503){59}{\rule{1.509pt}{0.121pt}}
\multiput(901.00,407.34)(50.868,-33.000){2}{\rule{0.755pt}{0.800pt}}
\multiput(955.00,374.09)(0.718,-0.503){67}{\rule{1.346pt}{0.121pt}}
\multiput(955.00,374.34)(50.206,-37.000){2}{\rule{0.673pt}{0.800pt}}
\multiput(1008.00,337.09)(0.694,-0.503){71}{\rule{1.308pt}{0.121pt}}
\multiput(1008.00,337.34)(51.286,-39.000){2}{\rule{0.654pt}{0.800pt}}
\multiput(1062.00,298.09)(0.663,-0.502){73}{\rule{1.260pt}{0.121pt}}
\multiput(1062.00,298.34)(50.385,-40.000){2}{\rule{0.630pt}{0.800pt}}
\multiput(1115.00,258.09)(0.694,-0.503){71}{\rule{1.308pt}{0.121pt}}
\multiput(1115.00,258.34)(51.286,-39.000){2}{\rule{0.654pt}{0.800pt}}
\multiput(1169.00,219.09)(0.718,-0.503){67}{\rule{1.346pt}{0.121pt}}
\multiput(1169.00,219.34)(50.206,-37.000){2}{\rule{0.673pt}{0.800pt}}
\multiput(1222.00,182.09)(0.823,-0.503){59}{\rule{1.509pt}{0.121pt}}
\multiput(1222.00,182.34)(50.868,-33.000){2}{\rule{0.755pt}{0.800pt}}
\multiput(1276.00,149.09)(0.922,-0.504){51}{\rule{1.662pt}{0.121pt}}
\multiput(1276.00,149.34)(49.550,-29.000){2}{\rule{0.831pt}{0.800pt}}
\sbox{\plotpoint}{\rule[-0.500pt]{1.000pt}{1.000pt}}%
\put(220,848){\usebox{\plotpoint}}
\multiput(220,848)(20.756,0.000){59}{\usebox{\plotpoint}}
\put(1436,848){\usebox{\plotpoint}}
\put(400,790){$5\sigma$}
\end{picture}

\vskip 2 true cm

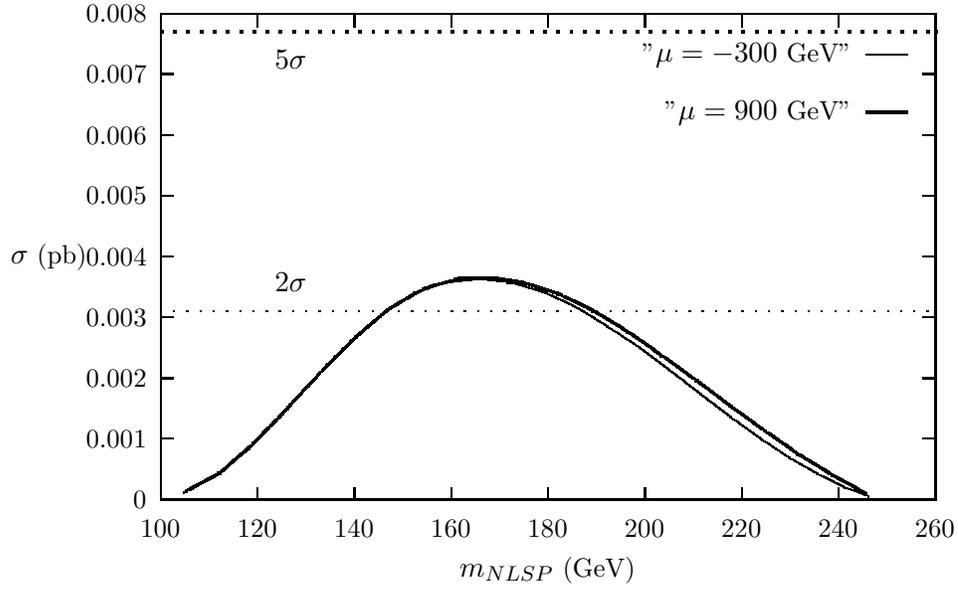
\centerline{\bf Fig. 2}

\noindent The cross section for $e^+e^-\longrightarrow \gamma+$
 missing energy
against the NLSP mass with unpolarized electron beams 
at $E_{cm}=500$ GeV. Two values of $\mu$ are used with $\tan\beta
=10$ and $M/\Lambda=2$. Different levels of significance correspond 
to an integrated luminosity of 30 fb$^{-1}$.

\newpage
\setlength{\unitlength}{0.240900pt}
\ifx\plotpoint\undefined\newsavebox{\plotpoint}\fi
\sbox{\plotpoint}{\rule[-0.200pt]{0.400pt}{0.400pt}}%
\begin{picture}(1500,900)(0,0)
\font\gnuplot=cmr10 at 10pt
\gnuplot
\sbox{\plotpoint}{\rule[-0.200pt]{0.400pt}{0.400pt}}%
\put(220.0,113.0){\rule[-0.200pt]{292.934pt}{0.400pt}}
\put(220.0,113.0){\rule[-0.200pt]{4.818pt}{0.400pt}}
\put(198,113){\makebox(0,0)[r]{0}}
\put(1416.0,113.0){\rule[-0.200pt]{4.818pt}{0.400pt}}
\put(220.0,189.0){\rule[-0.200pt]{4.818pt}{0.400pt}}
\put(198,189){\makebox(0,0)[r]{0.001}}
\put(1416.0,189.0){\rule[-0.200pt]{4.818pt}{0.400pt}}
\put(220.0,266.0){\rule[-0.200pt]{4.818pt}{0.400pt}}
\put(198,266){\makebox(0,0)[r]{0.002}}
\put(1416.0,266.0){\rule[-0.200pt]{4.818pt}{0.400pt}}
\put(220.0,342.0){\rule[-0.200pt]{4.818pt}{0.400pt}}
\put(198,342){\makebox(0,0)[r]{0.003}}
\put(1416.0,342.0){\rule[-0.200pt]{4.818pt}{0.400pt}}
\put(220.0,419.0){\rule[-0.200pt]{4.818pt}{0.400pt}}
\put(198,419){\makebox(0,0)[r]{0.004}}
\put(1416.0,419.0){\rule[-0.200pt]{4.818pt}{0.400pt}}
\put(220.0,495.0){\rule[-0.200pt]{4.818pt}{0.400pt}}
\put(198,495){\makebox(0,0)[r]{0.005}}
\put(1416.0,495.0){\rule[-0.200pt]{4.818pt}{0.400pt}}
\put(220.0,571.0){\rule[-0.200pt]{4.818pt}{0.400pt}}
\put(198,571){\makebox(0,0)[r]{0.006}}
\put(1416.0,571.0){\rule[-0.200pt]{4.818pt}{0.400pt}}
\put(220.0,648.0){\rule[-0.200pt]{4.818pt}{0.400pt}}
\put(198,648){\makebox(0,0)[r]{0.007}}
\put(1416.0,648.0){\rule[-0.200pt]{4.818pt}{0.400pt}}
\put(220.0,724.0){\rule[-0.200pt]{4.818pt}{0.400pt}}
\put(198,724){\makebox(0,0)[r]{0.008}}
\put(1416.0,724.0){\rule[-0.200pt]{4.818pt}{0.400pt}}
\put(220.0,801.0){\rule[-0.200pt]{4.818pt}{0.400pt}}
\put(198,801){\makebox(0,0)[r]{0.009}}
\put(1416.0,801.0){\rule[-0.200pt]{4.818pt}{0.400pt}}
\put(220.0,877.0){\rule[-0.200pt]{4.818pt}{0.400pt}}
\put(198,877){\makebox(0,0)[r]{0.01}}
\put(1416.0,877.0){\rule[-0.200pt]{4.818pt}{0.400pt}}
\put(220.0,113.0){\rule[-0.200pt]{0.400pt}{4.818pt}}
\put(220,68){\makebox(0,0){100}}
\put(220.0,857.0){\rule[-0.200pt]{0.400pt}{4.818pt}}
\put(372.0,113.0){\rule[-0.200pt]{0.400pt}{4.818pt}}
\put(372,68){\makebox(0,0){120}}
\put(372.0,857.0){\rule[-0.200pt]{0.400pt}{4.818pt}}
\put(524.0,113.0){\rule[-0.200pt]{0.400pt}{4.818pt}}
\put(524,68){\makebox(0,0){140}}
\put(524.0,857.0){\rule[-0.200pt]{0.400pt}{4.818pt}}
\put(676.0,113.0){\rule[-0.200pt]{0.400pt}{4.818pt}}
\put(676,68){\makebox(0,0){160}}
\put(676.0,857.0){\rule[-0.200pt]{0.400pt}{4.818pt}}
\put(828.0,113.0){\rule[-0.200pt]{0.400pt}{4.818pt}}
\put(828,68){\makebox(0,0){180}}
\put(828.0,857.0){\rule[-0.200pt]{0.400pt}{4.818pt}}
\put(980.0,113.0){\rule[-0.200pt]{0.400pt}{4.818pt}}
\put(980,68){\makebox(0,0){200}}
\put(980.0,857.0){\rule[-0.200pt]{0.400pt}{4.818pt}}
\put(1132.0,113.0){\rule[-0.200pt]{0.400pt}{4.818pt}}
\put(1132,68){\makebox(0,0){220}}
\put(1132.0,857.0){\rule[-0.200pt]{0.400pt}{4.818pt}}
\put(1284.0,113.0){\rule[-0.200pt]{0.400pt}{4.818pt}}
\put(1284,68){\makebox(0,0){240}}
\put(1284.0,857.0){\rule[-0.200pt]{0.400pt}{4.818pt}}
\put(1436.0,113.0){\rule[-0.200pt]{0.400pt}{4.818pt}}
\put(1436,68){\makebox(0,0){260}}
\put(1436.0,857.0){\rule[-0.200pt]{0.400pt}{4.818pt}}
\put(220.0,113.0){\rule[-0.200pt]{292.934pt}{0.400pt}}
\put(1436.0,113.0){\rule[-0.200pt]{0.400pt}{184.048pt}}
\put(220.0,877.0){\rule[-0.200pt]{292.934pt}{0.400pt}}
\put(35,495){\makebox(0,0){$\sigma$ (pb)}}
\put(828,23){\makebox(0,0){$m_{NLSP}$ (GeV)}}
\put(220.0,113.0){\rule[-0.200pt]{0.400pt}{184.048pt}}
\put(1306,812){\makebox(0,0)[r]{"$\mu=-300$ GeV"}}
\put(1328.0,812.0){\rule[-0.200pt]{15.899pt}{0.400pt}}
\put(256,131){\usebox{\plotpoint}}
\multiput(256.00,131.58)(0.619,0.498){81}{\rule{0.595pt}{0.120pt}}
\multiput(256.00,130.17)(50.765,42.000){2}{\rule{0.298pt}{0.400pt}}
\multiput(308.58,173.00)(0.498,0.644){101}{\rule{0.120pt}{0.615pt}}
\multiput(307.17,173.00)(52.000,65.723){2}{\rule{0.400pt}{0.308pt}}
\multiput(360.58,240.00)(0.498,0.809){101}{\rule{0.120pt}{0.746pt}}
\multiput(359.17,240.00)(52.000,82.451){2}{\rule{0.400pt}{0.373pt}}
\multiput(412.58,324.00)(0.498,0.867){101}{\rule{0.120pt}{0.792pt}}
\multiput(411.17,324.00)(52.000,88.356){2}{\rule{0.400pt}{0.396pt}}
\multiput(464.58,414.00)(0.498,0.825){99}{\rule{0.120pt}{0.759pt}}
\multiput(463.17,414.00)(51.000,82.425){2}{\rule{0.400pt}{0.379pt}}
\multiput(515.58,498.00)(0.498,0.703){101}{\rule{0.120pt}{0.662pt}}
\multiput(514.17,498.00)(52.000,71.627){2}{\rule{0.400pt}{0.331pt}}
\multiput(567.58,571.00)(0.498,0.529){99}{\rule{0.120pt}{0.524pt}}
\multiput(566.17,571.00)(51.000,52.913){2}{\rule{0.400pt}{0.262pt}}
\multiput(618.00,625.58)(0.968,0.497){51}{\rule{0.870pt}{0.120pt}}
\multiput(618.00,624.17)(50.194,27.000){2}{\rule{0.435pt}{0.400pt}}
\multiput(670.00,652.59)(2.941,0.489){15}{\rule{2.367pt}{0.118pt}}
\multiput(670.00,651.17)(46.088,9.000){2}{\rule{1.183pt}{0.400pt}}
\multiput(721.00,659.92)(2.580,-0.491){17}{\rule{2.100pt}{0.118pt}}
\multiput(721.00,660.17)(45.641,-10.000){2}{\rule{1.050pt}{0.400pt}}
\multiput(771.00,649.92)(1.117,-0.496){43}{\rule{0.987pt}{0.120pt}}
\multiput(771.00,650.17)(48.952,-23.000){2}{\rule{0.493pt}{0.400pt}}
\multiput(822.00,626.92)(0.695,-0.498){69}{\rule{0.656pt}{0.120pt}}
\multiput(822.00,627.17)(48.639,-36.000){2}{\rule{0.328pt}{0.400pt}}
\multiput(872.00,590.92)(0.520,-0.498){93}{\rule{0.517pt}{0.120pt}}
\multiput(872.00,591.17)(48.928,-48.000){2}{\rule{0.258pt}{0.400pt}}
\multiput(922.58,541.82)(0.498,-0.530){95}{\rule{0.120pt}{0.524pt}}
\multiput(921.17,542.91)(49.000,-50.911){2}{\rule{0.400pt}{0.262pt}}
\multiput(971.58,489.65)(0.498,-0.581){95}{\rule{0.120pt}{0.565pt}}
\multiput(970.17,490.83)(49.000,-55.827){2}{\rule{0.400pt}{0.283pt}}
\multiput(1020.58,432.54)(0.498,-0.615){93}{\rule{0.120pt}{0.592pt}}
\multiput(1019.17,433.77)(48.000,-57.772){2}{\rule{0.400pt}{0.296pt}}
\multiput(1068.58,373.61)(0.498,-0.596){91}{\rule{0.120pt}{0.577pt}}
\multiput(1067.17,374.80)(47.000,-54.803){2}{\rule{0.400pt}{0.288pt}}
\multiput(1115.58,317.68)(0.498,-0.574){91}{\rule{0.120pt}{0.560pt}}
\multiput(1114.17,318.84)(47.000,-52.839){2}{\rule{0.400pt}{0.280pt}}
\multiput(1162.58,263.81)(0.498,-0.533){87}{\rule{0.120pt}{0.527pt}}
\multiput(1161.17,264.91)(45.000,-46.907){2}{\rule{0.400pt}{0.263pt}}
\multiput(1207.00,216.92)(0.549,-0.498){77}{\rule{0.540pt}{0.120pt}}
\multiput(1207.00,217.17)(42.879,-40.000){2}{\rule{0.270pt}{0.400pt}}
\put(220,364){\usebox{\plotpoint}}
\multiput(220,364)(20.756,0.000){59}{\usebox{\plotpoint}}
\put(1436,364){\usebox{\plotpoint}}
\put(275,324){$5\sigma$}
\sbox{\plotpoint}{\rule[-0.400pt]{0.800pt}{0.800pt}}%
\put(1306,722){\makebox(0,0)[r]{"$\mu=900$ GeV"}}
\put(1328.0,722.0){\rule[-0.400pt]{15.899pt}{0.800pt}}
\put(259,133){\usebox{\plotpoint}}
\multiput(259.00,134.41)(0.587,0.502){85}{\rule{1.139pt}{0.121pt}}
\multiput(259.00,131.34)(51.636,46.000){2}{\rule{0.570pt}{0.800pt}}
\multiput(314.41,179.00)(0.502,0.671){99}{\rule{0.121pt}{1.272pt}}
\multiput(311.34,179.00)(53.000,68.361){2}{\rule{0.800pt}{0.636pt}}
\multiput(367.41,250.00)(0.502,0.808){101}{\rule{0.121pt}{1.489pt}}
\multiput(364.34,250.00)(54.000,83.910){2}{\rule{0.800pt}{0.744pt}}
\multiput(421.41,337.00)(0.502,0.871){99}{\rule{0.121pt}{1.589pt}}
\multiput(418.34,337.00)(53.000,88.703){2}{\rule{0.800pt}{0.794pt}}
\multiput(474.41,429.00)(0.502,0.789){101}{\rule{0.121pt}{1.459pt}}
\multiput(471.34,429.00)(54.000,81.971){2}{\rule{0.800pt}{0.730pt}}
\multiput(528.41,514.00)(0.502,0.680){99}{\rule{0.121pt}{1.287pt}}
\multiput(525.34,514.00)(53.000,69.329){2}{\rule{0.800pt}{0.643pt}}
\multiput(580.00,587.41)(0.550,0.502){91}{\rule{1.082pt}{0.121pt}}
\multiput(580.00,584.34)(51.755,49.000){2}{\rule{0.541pt}{0.800pt}}
\multiput(634.00,636.41)(1.121,0.504){41}{\rule{1.967pt}{0.122pt}}
\multiput(634.00,633.34)(48.918,24.000){2}{\rule{0.983pt}{0.800pt}}
\put(687,658.84){\rule{13.009pt}{0.800pt}}
\multiput(687.00,657.34)(27.000,3.000){2}{\rule{6.504pt}{0.800pt}}
\multiput(741.00,660.08)(2.146,-0.509){19}{\rule{3.462pt}{0.123pt}}
\multiput(741.00,660.34)(45.815,-13.000){2}{\rule{1.731pt}{0.800pt}}
\multiput(794.00,647.09)(0.974,-0.504){49}{\rule{1.743pt}{0.121pt}}
\multiput(794.00,647.34)(50.383,-28.000){2}{\rule{0.871pt}{0.800pt}}
\multiput(848.00,619.09)(0.631,-0.502){77}{\rule{1.210pt}{0.121pt}}
\multiput(848.00,619.34)(50.490,-42.000){2}{\rule{0.605pt}{0.800pt}}
\multiput(901.00,577.09)(0.528,-0.502){95}{\rule{1.047pt}{0.121pt}}
\multiput(901.00,577.34)(51.827,-51.000){2}{\rule{0.524pt}{0.800pt}}
\multiput(956.41,523.60)(0.502,-0.537){99}{\rule{0.121pt}{1.060pt}}
\multiput(953.34,525.80)(53.000,-54.799){2}{\rule{0.800pt}{0.530pt}}
\multiput(1009.41,466.30)(0.502,-0.583){101}{\rule{0.121pt}{1.133pt}}
\multiput(1006.34,468.65)(54.000,-60.648){2}{\rule{0.800pt}{0.567pt}}
\multiput(1063.41,403.28)(0.502,-0.585){99}{\rule{0.121pt}{1.136pt}}
\multiput(1060.34,405.64)(53.000,-59.642){2}{\rule{0.800pt}{0.568pt}}
\multiput(1116.41,341.30)(0.502,-0.583){101}{\rule{0.121pt}{1.133pt}}
\multiput(1113.34,343.65)(54.000,-60.648){2}{\rule{0.800pt}{0.567pt}}
\multiput(1170.41,278.54)(0.502,-0.546){99}{\rule{0.121pt}{1.075pt}}
\multiput(1167.34,280.77)(53.000,-55.768){2}{\rule{0.800pt}{0.538pt}}
\multiput(1222.00,223.09)(0.518,-0.502){97}{\rule{1.031pt}{0.121pt}}
\multiput(1222.00,223.34)(51.861,-52.000){2}{\rule{0.515pt}{0.800pt}}
\multiput(1276.00,171.09)(0.589,-0.502){83}{\rule{1.142pt}{0.121pt}}
\multiput(1276.00,171.34)(50.629,-45.000){2}{\rule{0.571pt}{0.800pt}}
\sbox{\plotpoint}{\rule[-0.500pt]{1.000pt}{1.000pt}}%
\put(220,614){\usebox{\plotpoint}}
\multiput(220,614)(20.756,0.000){59}{\usebox{\plotpoint}}
\put(1436,614){\usebox{\plotpoint}}
\put(275,574){$10\sigma$}
\sbox{\plotpoint}{\rule[-0.600pt]{1.200pt}{1.200pt}}%
\end{picture}

\vskip 2 true cm

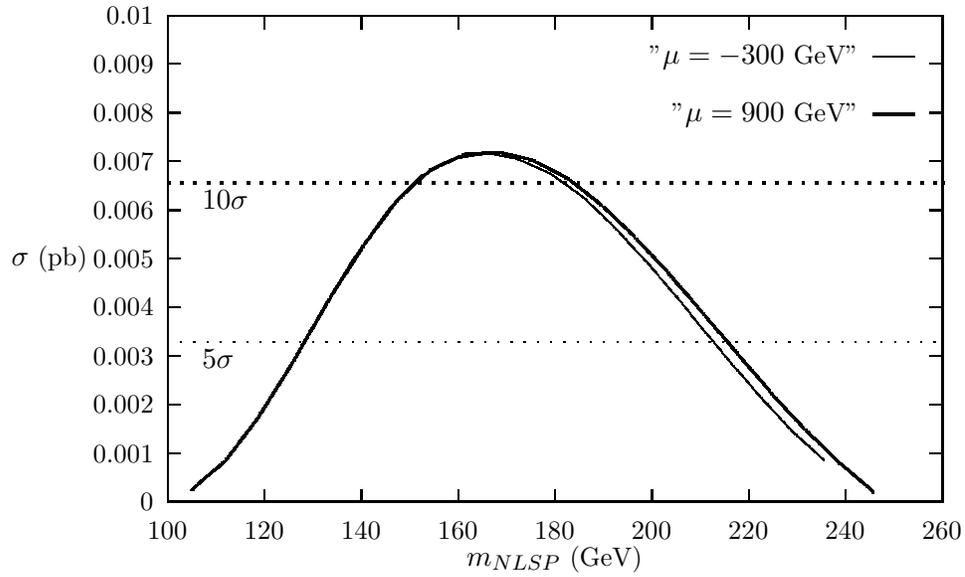
\centerline{\bf Fig. 3}

\noindent Same as in Figure 2, but with 100\% right-polarized
electron beams. Also, different levels of significance correspond to 
an integrated luminosity of 10 fb$^{-1}$.

\newpage
\setlength{\unitlength}{0.240900pt}
\ifx\plotpoint\undefined\newsavebox{\plotpoint}\fi
\sbox{\plotpoint}{\rule[-0.200pt]{0.400pt}{0.400pt}}%
\begin{picture}(1500,900)(0,0)
\font\gnuplot=cmr10 at 10pt
\gnuplot
\sbox{\plotpoint}{\rule[-0.200pt]{0.400pt}{0.400pt}}%
\put(220.0,113.0){\rule[-0.200pt]{4.818pt}{0.400pt}}
\put(198,113){\makebox(0,0)[r]{0.02}}
\put(1416.0,113.0){\rule[-0.200pt]{4.818pt}{0.400pt}}
\put(220.0,222.0){\rule[-0.200pt]{4.818pt}{0.400pt}}
\put(198,222){\makebox(0,0)[r]{0.03}}
\put(1416.0,222.0){\rule[-0.200pt]{4.818pt}{0.400pt}}
\put(220.0,331.0){\rule[-0.200pt]{4.818pt}{0.400pt}}
\put(198,331){\makebox(0,0)[r]{0.04}}
\put(1416.0,331.0){\rule[-0.200pt]{4.818pt}{0.400pt}}
\put(220.0,440.0){\rule[-0.200pt]{4.818pt}{0.400pt}}
\put(198,440){\makebox(0,0)[r]{0.05}}
\put(1416.0,440.0){\rule[-0.200pt]{4.818pt}{0.400pt}}
\put(220.0,550.0){\rule[-0.200pt]{4.818pt}{0.400pt}}
\put(198,550){\makebox(0,0)[r]{0.06}}
\put(1416.0,550.0){\rule[-0.200pt]{4.818pt}{0.400pt}}
\put(220.0,659.0){\rule[-0.200pt]{4.818pt}{0.400pt}}
\put(198,659){\makebox(0,0)[r]{0.07}}
\put(1416.0,659.0){\rule[-0.200pt]{4.818pt}{0.400pt}}
\put(220.0,768.0){\rule[-0.200pt]{4.818pt}{0.400pt}}
\put(198,768){\makebox(0,0)[r]{0.08}}
\put(1416.0,768.0){\rule[-0.200pt]{4.818pt}{0.400pt}}
\put(220.0,877.0){\rule[-0.200pt]{4.818pt}{0.400pt}}
\put(198,877){\makebox(0,0)[r]{0.09}}
\put(1416.0,877.0){\rule[-0.200pt]{4.818pt}{0.400pt}}
\put(314.0,113.0){\rule[-0.200pt]{0.400pt}{4.818pt}}
\put(314,68){\makebox(0,0){400}}
\put(314.0,857.0){\rule[-0.200pt]{0.400pt}{4.818pt}}
\put(501.0,113.0){\rule[-0.200pt]{0.400pt}{4.818pt}}
\put(501,68){\makebox(0,0){500}}
\put(501.0,857.0){\rule[-0.200pt]{0.400pt}{4.818pt}}
\put(688.0,113.0){\rule[-0.200pt]{0.400pt}{4.818pt}}
\put(688,68){\makebox(0,0){600}}
\put(688.0,857.0){\rule[-0.200pt]{0.400pt}{4.818pt}}
\put(875.0,113.0){\rule[-0.200pt]{0.400pt}{4.818pt}}
\put(875,68){\makebox(0,0){700}}
\put(875.0,857.0){\rule[-0.200pt]{0.400pt}{4.818pt}}
\put(1062.0,113.0){\rule[-0.200pt]{0.400pt}{4.818pt}}
\put(1062,68){\makebox(0,0){800}}
\put(1062.0,857.0){\rule[-0.200pt]{0.400pt}{4.818pt}}
\put(1249.0,113.0){\rule[-0.200pt]{0.400pt}{4.818pt}}
\put(1249,68){\makebox(0,0){900}}
\put(1249.0,857.0){\rule[-0.200pt]{0.400pt}{4.818pt}}
\put(1436.0,113.0){\rule[-0.200pt]{0.400pt}{4.818pt}}
\put(1436,68){\makebox(0,0){1000}}
\put(1436.0,857.0){\rule[-0.200pt]{0.400pt}{4.818pt}}
\put(220.0,113.0){\rule[-0.200pt]{292.934pt}{0.400pt}}
\put(1436.0,113.0){\rule[-0.200pt]{0.400pt}{184.048pt}}
\put(220.0,877.0){\rule[-0.200pt]{292.934pt}{0.400pt}}
\put(45,495){\makebox(0,0){$\sigma$ (pb)}}
\put(828,23){\makebox(0,0){$\sqrt{s}$ (GeV)}}
\put(220.0,113.0){\rule[-0.200pt]{0.400pt}{184.048pt}}
\put(1306,812){\makebox(0,0)[r]{"$\tilde\nu + $ LSP $+\gamma$"}}
\put(1328.0,812.0){\rule[-0.200pt]{15.899pt}{0.400pt}}
\put(220,867){\usebox{\plotpoint}}
\put(220,865.67){\rule{22.645pt}{0.400pt}}
\multiput(220.00,866.17)(47.000,-1.000){2}{\rule{11.322pt}{0.400pt}}
\multiput(314.00,864.92)(0.951,-0.498){95}{\rule{0.859pt}{0.120pt}}
\multiput(314.00,865.17)(91.217,-49.000){2}{\rule{0.430pt}{0.400pt}}
\multiput(407.00,815.92)(0.812,-0.499){113}{\rule{0.748pt}{0.120pt}}
\multiput(407.00,816.17)(92.447,-58.000){2}{\rule{0.374pt}{0.400pt}}
\multiput(501.00,757.92)(0.739,-0.499){123}{\rule{0.690pt}{0.120pt}}
\multiput(501.00,758.17)(91.567,-63.000){2}{\rule{0.345pt}{0.400pt}}
\multiput(594.00,694.92)(0.724,-0.499){127}{\rule{0.678pt}{0.120pt}}
\multiput(594.00,695.17)(92.592,-65.000){2}{\rule{0.339pt}{0.400pt}}
\multiput(688.00,629.92)(0.914,-0.498){99}{\rule{0.829pt}{0.120pt}}
\multiput(688.00,630.17)(91.279,-51.000){2}{\rule{0.415pt}{0.400pt}}
\multiput(781.00,578.92)(0.841,-0.499){109}{\rule{0.771pt}{0.120pt}}
\multiput(781.00,579.17)(92.399,-56.000){2}{\rule{0.386pt}{0.400pt}}
\multiput(875.00,522.92)(0.932,-0.498){97}{\rule{0.844pt}{0.120pt}}
\multiput(875.00,523.17)(91.248,-50.000){2}{\rule{0.422pt}{0.400pt}}
\multiput(968.00,472.92)(1.180,-0.498){77}{\rule{1.040pt}{0.120pt}}
\multiput(968.00,473.17)(91.841,-40.000){2}{\rule{0.520pt}{0.400pt}}
\multiput(1062.00,432.92)(1.139,-0.498){79}{\rule{1.007pt}{0.120pt}}
\multiput(1062.00,433.17)(90.909,-41.000){2}{\rule{0.504pt}{0.400pt}}
\multiput(1155.00,391.92)(1.277,-0.498){71}{\rule{1.116pt}{0.120pt}}
\multiput(1155.00,392.17)(91.683,-37.000){2}{\rule{0.558pt}{0.400pt}}
\multiput(1249.00,354.92)(1.418,-0.497){63}{\rule{1.227pt}{0.120pt}}
\multiput(1249.00,355.17)(90.453,-33.000){2}{\rule{0.614pt}{0.400pt}}
\multiput(1342.00,321.92)(1.693,-0.497){53}{\rule{1.443pt}{0.120pt}}
\multiput(1342.00,322.17)(91.005,-28.000){2}{\rule{0.721pt}{0.400pt}}
\put(1306,767){\makebox(0,0)[r]{"Neutralino LSP $+\gamma$"}}
\multiput(1328,767)(20.756,0.000){4}{\usebox{\plotpoint}}
\put(1394,767){\usebox{\plotpoint}}
\put(220,286){\usebox{\plotpoint}}
\multiput(220,286)(16.987,11.927){6}{\usebox{\plotpoint}}
\multiput(314,352)(19.874,5.984){5}{\usebox{\plotpoint}}
\multiput(407,380)(20.698,1.541){4}{\usebox{\plotpoint}}
\multiput(501,387)(20.637,-2.219){5}{\usebox{\plotpoint}}
\multiput(594,377)(20.385,-3.904){4}{\usebox{\plotpoint}}
\multiput(688,359)(20.097,-5.186){5}{\usebox{\plotpoint}}
\multiput(781,335)(20.256,-4.525){5}{\usebox{\plotpoint}}
\multiput(875,314)(20.198,-4.778){4}{\usebox{\plotpoint}}
\multiput(968,292)(20.110,-5.135){5}{\usebox{\plotpoint}}
\multiput(1062,268)(20.246,-4.572){5}{\usebox{\plotpoint}}
\multiput(1155,247)(20.256,-4.525){4}{\usebox{\plotpoint}}
\multiput(1249,226)(20.335,-4.155){5}{\usebox{\plotpoint}}
\multiput(1342,207)(20.754,-0.221){4}{\usebox{\plotpoint}}
\put(1436,206){\usebox{\plotpoint}}
\sbox{\plotpoint}{\rule[-0.400pt]{0.800pt}{0.800pt}}%
\put(1306,722){\makebox(0,0)[r]{"Total VLSP $+\gamma$"}}
\put(1328.0,722.0){\rule[-0.400pt]{15.899pt}{0.800pt}}
\put(220,713){\usebox{\plotpoint}}
\multiput(220.00,714.41)(1.921,0.504){43}{\rule{3.208pt}{0.121pt}}
\multiput(220.00,711.34)(87.342,25.000){2}{\rule{1.604pt}{0.800pt}}
\multiput(314.00,736.09)(1.983,-0.504){41}{\rule{3.300pt}{0.122pt}}
\multiput(314.00,736.34)(86.151,-24.000){2}{\rule{1.650pt}{0.800pt}}
\multiput(407.00,712.09)(1.320,-0.503){65}{\rule{2.289pt}{0.121pt}}
\multiput(407.00,712.34)(89.249,-36.000){2}{\rule{1.144pt}{0.800pt}}
\multiput(501.00,676.09)(0.974,-0.502){89}{\rule{1.750pt}{0.121pt}}
\multiput(501.00,676.34)(89.368,-48.000){2}{\rule{0.875pt}{0.800pt}}
\multiput(594.00,628.09)(0.908,-0.502){97}{\rule{1.646pt}{0.121pt}}
\multiput(594.00,628.34)(90.583,-52.000){2}{\rule{0.823pt}{0.800pt}}
\multiput(688.00,576.09)(0.995,-0.502){87}{\rule{1.783pt}{0.121pt}}
\multiput(688.00,576.34)(89.299,-47.000){2}{\rule{0.891pt}{0.800pt}}
\multiput(781.00,529.09)(1.028,-0.502){85}{\rule{1.835pt}{0.121pt}}
\multiput(781.00,529.34)(90.192,-46.000){2}{\rule{0.917pt}{0.800pt}}
\multiput(875.00,483.09)(1.064,-0.502){81}{\rule{1.891pt}{0.121pt}}
\multiput(875.00,483.34)(89.075,-44.000){2}{\rule{0.945pt}{0.800pt}}
\multiput(968.00,439.09)(1.320,-0.503){65}{\rule{2.289pt}{0.121pt}}
\multiput(968.00,439.34)(89.249,-36.000){2}{\rule{1.144pt}{0.800pt}}
\multiput(1062.00,403.09)(1.344,-0.503){63}{\rule{2.326pt}{0.121pt}}
\multiput(1062.00,403.34)(88.173,-35.000){2}{\rule{1.163pt}{0.800pt}}
\multiput(1155.00,368.09)(1.400,-0.503){61}{\rule{2.412pt}{0.121pt}}
\multiput(1155.00,368.34)(88.994,-34.000){2}{\rule{1.206pt}{0.800pt}}
\multiput(1249.00,334.09)(1.575,-0.503){53}{\rule{2.680pt}{0.121pt}}
\multiput(1249.00,334.34)(87.438,-30.000){2}{\rule{1.340pt}{0.800pt}}
\multiput(1342.00,304.09)(1.709,-0.504){49}{\rule{2.886pt}{0.121pt}}
\multiput(1342.00,304.34)(88.011,-28.000){2}{\rule{1.443pt}{0.800pt}}
\end{picture}

\vskip 2 true cm

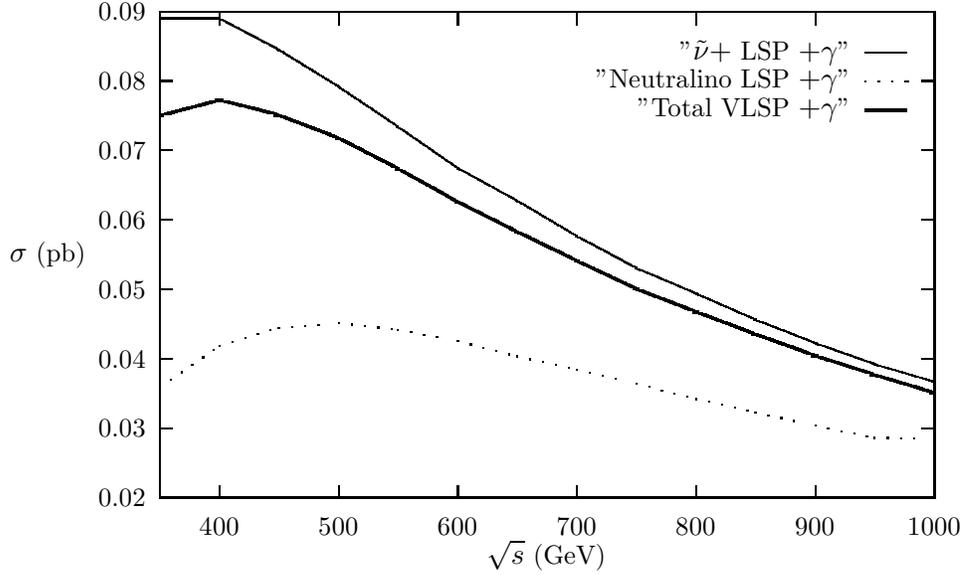
\centerline{\bf Fig. 4}

\noindent Single photon production as a function 
of total center-of-mass energy
$\sqrt{s}$. The three lines are drawn with different set of parameters:
(i) the contribution with $\tilde\nu$ and LSP (lightest neutralino)
at $m_{\chi^0_1}=71$ GeV and $m_{\tilde\nu}=100$ GeV; 
(ii) the contribution with Virtual LSPs ($\tilde\nu$ and two lightest
neutralinos)
at $m_{\chi^0_1}=71$ GeV and $m_{\tilde\nu}=120$ GeV; 
(iii) the contribution with LSP only
at $m_{\chi^0_1}=100$ GeV and $m_{\tilde\nu}=150$ GeV. Gauge coupling
unification is assumed.

\newpage
\setlength{\unitlength}{0.240900pt}
\ifx\plotpoint\undefined\newsavebox{\plotpoint}\fi
\sbox{\plotpoint}{\rule[-0.200pt]{0.400pt}{0.400pt}}%
\begin{picture}(1500,900)(0,0)
\font\gnuplot=cmr10 at 10pt
\gnuplot
\sbox{\plotpoint}{\rule[-0.200pt]{0.400pt}{0.400pt}}%
\put(220.0,113.0){\rule[-0.200pt]{0.400pt}{184.048pt}}
\put(220.0,113.0){\rule[-0.200pt]{4.818pt}{0.400pt}}
\put(198,113){\makebox(0,0)[r]{1e-05}}
\put(1416.0,113.0){\rule[-0.200pt]{4.818pt}{0.400pt}}
\put(220.0,190.0){\rule[-0.200pt]{2.409pt}{0.400pt}}
\put(1426.0,190.0){\rule[-0.200pt]{2.409pt}{0.400pt}}
\put(220.0,235.0){\rule[-0.200pt]{2.409pt}{0.400pt}}
\put(1426.0,235.0){\rule[-0.200pt]{2.409pt}{0.400pt}}
\put(220.0,266.0){\rule[-0.200pt]{2.409pt}{0.400pt}}
\put(1426.0,266.0){\rule[-0.200pt]{2.409pt}{0.400pt}}
\put(220.0,291.0){\rule[-0.200pt]{2.409pt}{0.400pt}}
\put(1426.0,291.0){\rule[-0.200pt]{2.409pt}{0.400pt}}
\put(220.0,311.0){\rule[-0.200pt]{2.409pt}{0.400pt}}
\put(1426.0,311.0){\rule[-0.200pt]{2.409pt}{0.400pt}}
\put(220.0,328.0){\rule[-0.200pt]{2.409pt}{0.400pt}}
\put(1426.0,328.0){\rule[-0.200pt]{2.409pt}{0.400pt}}
\put(220.0,343.0){\rule[-0.200pt]{2.409pt}{0.400pt}}
\put(1426.0,343.0){\rule[-0.200pt]{2.409pt}{0.400pt}}
\put(220.0,356.0){\rule[-0.200pt]{2.409pt}{0.400pt}}
\put(1426.0,356.0){\rule[-0.200pt]{2.409pt}{0.400pt}}
\put(220.0,368.0){\rule[-0.200pt]{4.818pt}{0.400pt}}
\put(198,368){\makebox(0,0)[r]{0.0001}}
\put(1416.0,368.0){\rule[-0.200pt]{4.818pt}{0.400pt}}
\put(220.0,444.0){\rule[-0.200pt]{2.409pt}{0.400pt}}
\put(1426.0,444.0){\rule[-0.200pt]{2.409pt}{0.400pt}}
\put(220.0,489.0){\rule[-0.200pt]{2.409pt}{0.400pt}}
\put(1426.0,489.0){\rule[-0.200pt]{2.409pt}{0.400pt}}
\put(220.0,521.0){\rule[-0.200pt]{2.409pt}{0.400pt}}
\put(1426.0,521.0){\rule[-0.200pt]{2.409pt}{0.400pt}}
\put(220.0,546.0){\rule[-0.200pt]{2.409pt}{0.400pt}}
\put(1426.0,546.0){\rule[-0.200pt]{2.409pt}{0.400pt}}
\put(220.0,566.0){\rule[-0.200pt]{2.409pt}{0.400pt}}
\put(1426.0,566.0){\rule[-0.200pt]{2.409pt}{0.400pt}}
\put(220.0,583.0){\rule[-0.200pt]{2.409pt}{0.400pt}}
\put(1426.0,583.0){\rule[-0.200pt]{2.409pt}{0.400pt}}
\put(220.0,598.0){\rule[-0.200pt]{2.409pt}{0.400pt}}
\put(1426.0,598.0){\rule[-0.200pt]{2.409pt}{0.400pt}}
\put(220.0,611.0){\rule[-0.200pt]{2.409pt}{0.400pt}}
\put(1426.0,611.0){\rule[-0.200pt]{2.409pt}{0.400pt}}
\put(220.0,622.0){\rule[-0.200pt]{4.818pt}{0.400pt}}
\put(198,622){\makebox(0,0)[r]{0.001}}
\put(1416.0,622.0){\rule[-0.200pt]{4.818pt}{0.400pt}}
\put(220.0,699.0){\rule[-0.200pt]{2.409pt}{0.400pt}}
\put(1426.0,699.0){\rule[-0.200pt]{2.409pt}{0.400pt}}
\put(220.0,744.0){\rule[-0.200pt]{2.409pt}{0.400pt}}
\put(1426.0,744.0){\rule[-0.200pt]{2.409pt}{0.400pt}}
\put(220.0,776.0){\rule[-0.200pt]{2.409pt}{0.400pt}}
\put(1426.0,776.0){\rule[-0.200pt]{2.409pt}{0.400pt}}
\put(220.0,800.0){\rule[-0.200pt]{2.409pt}{0.400pt}}
\put(1426.0,800.0){\rule[-0.200pt]{2.409pt}{0.400pt}}
\put(220.0,821.0){\rule[-0.200pt]{2.409pt}{0.400pt}}
\put(1426.0,821.0){\rule[-0.200pt]{2.409pt}{0.400pt}}
\put(220.0,838.0){\rule[-0.200pt]{2.409pt}{0.400pt}}
\put(1426.0,838.0){\rule[-0.200pt]{2.409pt}{0.400pt}}
\put(220.0,852.0){\rule[-0.200pt]{2.409pt}{0.400pt}}
\put(1426.0,852.0){\rule[-0.200pt]{2.409pt}{0.400pt}}
\put(220.0,865.0){\rule[-0.200pt]{2.409pt}{0.400pt}}
\put(1426.0,865.0){\rule[-0.200pt]{2.409pt}{0.400pt}}
\put(220.0,877.0){\rule[-0.200pt]{4.818pt}{0.400pt}}
\put(198,877){\makebox(0,0)[r]{0.01}}
\put(1416.0,877.0){\rule[-0.200pt]{4.818pt}{0.400pt}}
\put(220.0,113.0){\rule[-0.200pt]{0.400pt}{4.818pt}}
\put(220,68){\makebox(0,0){0}}
\put(220.0,857.0){\rule[-0.200pt]{0.400pt}{4.818pt}}
\put(463.0,113.0){\rule[-0.200pt]{0.400pt}{4.818pt}}
\put(463,68){\makebox(0,0){50}}
\put(463.0,857.0){\rule[-0.200pt]{0.400pt}{4.818pt}}
\put(706.0,113.0){\rule[-0.200pt]{0.400pt}{4.818pt}}
\put(706,68){\makebox(0,0){100}}
\put(706.0,857.0){\rule[-0.200pt]{0.400pt}{4.818pt}}
\put(950.0,113.0){\rule[-0.200pt]{0.400pt}{4.818pt}}
\put(950,68){\makebox(0,0){150}}
\put(950.0,857.0){\rule[-0.200pt]{0.400pt}{4.818pt}}
\put(1193.0,113.0){\rule[-0.200pt]{0.400pt}{4.818pt}}
\put(1193,68){\makebox(0,0){200}}
\put(1193.0,857.0){\rule[-0.200pt]{0.400pt}{4.818pt}}
\put(1436.0,113.0){\rule[-0.200pt]{0.400pt}{4.818pt}}
\put(1436,68){\makebox(0,0){250}}
\put(1436.0,857.0){\rule[-0.200pt]{0.400pt}{4.818pt}}
\put(220.0,113.0){\rule[-0.200pt]{292.934pt}{0.400pt}}
\put(1436.0,113.0){\rule[-0.200pt]{0.400pt}{184.048pt}}
\put(220.0,877.0){\rule[-0.200pt]{292.934pt}{0.400pt}}
\put(45,495){\makebox(0,0){$d\sigma$ (pb)}}
\put(828,23){\makebox(0,0){$E_{\gamma}$ (GeV)}}
\put(220.0,113.0){\rule[-0.200pt]{0.400pt}{184.048pt}}
\put(1306,812){\makebox(0,0)[r]{"GMSB"}}
\put(1328.0,812.0){\rule[-0.200pt]{15.899pt}{0.400pt}}
\multiput(352.58,244.00)(0.496,5.465){37}{\rule{0.119pt}{4.400pt}}
\multiput(351.17,244.00)(20.000,205.868){2}{\rule{0.400pt}{2.200pt}}
\multiput(372.00,457.93)(1.286,-0.488){13}{\rule{1.100pt}{0.117pt}}
\multiput(372.00,458.17)(17.717,-8.000){2}{\rule{0.550pt}{0.400pt}}
\multiput(392.00,449.93)(2.269,-0.477){7}{\rule{1.780pt}{0.115pt}}
\multiput(392.00,450.17)(17.306,-5.000){2}{\rule{0.890pt}{0.400pt}}
\put(413,444.17){\rule{4.100pt}{0.400pt}}
\multiput(413.00,445.17)(11.490,-2.000){2}{\rule{2.050pt}{0.400pt}}
\put(352.0,113.0){\rule[-0.200pt]{0.400pt}{31.558pt}}
\put(453,442.67){\rule{4.818pt}{0.400pt}}
\multiput(453.00,443.17)(10.000,-1.000){2}{\rule{2.409pt}{0.400pt}}
\put(473,442.67){\rule{5.059pt}{0.400pt}}
\multiput(473.00,442.17)(10.500,1.000){2}{\rule{2.529pt}{0.400pt}}
\put(433.0,444.0){\rule[-0.200pt]{4.818pt}{0.400pt}}
\put(514,443.67){\rule{4.818pt}{0.400pt}}
\multiput(514.00,443.17)(10.000,1.000){2}{\rule{2.409pt}{0.400pt}}
\put(534,444.67){\rule{4.818pt}{0.400pt}}
\multiput(534.00,444.17)(10.000,1.000){2}{\rule{2.409pt}{0.400pt}}
\put(554,445.67){\rule{5.059pt}{0.400pt}}
\multiput(554.00,445.17)(10.500,1.000){2}{\rule{2.529pt}{0.400pt}}
\put(575,446.67){\rule{4.818pt}{0.400pt}}
\multiput(575.00,446.17)(10.000,1.000){2}{\rule{2.409pt}{0.400pt}}
\put(595,447.67){\rule{4.818pt}{0.400pt}}
\multiput(595.00,447.17)(10.000,1.000){2}{\rule{2.409pt}{0.400pt}}
\put(615,448.67){\rule{4.818pt}{0.400pt}}
\multiput(615.00,448.17)(10.000,1.000){2}{\rule{2.409pt}{0.400pt}}
\put(494.0,444.0){\rule[-0.200pt]{4.818pt}{0.400pt}}
\put(656,450.17){\rule{4.100pt}{0.400pt}}
\multiput(656.00,449.17)(11.490,2.000){2}{\rule{2.050pt}{0.400pt}}
\put(635.0,450.0){\rule[-0.200pt]{5.059pt}{0.400pt}}
\put(696,451.67){\rule{5.059pt}{0.400pt}}
\multiput(696.00,451.17)(10.500,1.000){2}{\rule{2.529pt}{0.400pt}}
\put(717,452.67){\rule{4.818pt}{0.400pt}}
\multiput(717.00,452.17)(10.000,1.000){2}{\rule{2.409pt}{0.400pt}}
\put(737,453.67){\rule{4.818pt}{0.400pt}}
\multiput(737.00,453.17)(10.000,1.000){2}{\rule{2.409pt}{0.400pt}}
\put(757,454.67){\rule{4.818pt}{0.400pt}}
\multiput(757.00,454.17)(10.000,1.000){2}{\rule{2.409pt}{0.400pt}}
\put(676.0,452.0){\rule[-0.200pt]{4.818pt}{0.400pt}}
\put(798,455.67){\rule{4.818pt}{0.400pt}}
\multiput(798.00,455.17)(10.000,1.000){2}{\rule{2.409pt}{0.400pt}}
\put(818,456.67){\rule{4.818pt}{0.400pt}}
\multiput(818.00,456.17)(10.000,1.000){2}{\rule{2.409pt}{0.400pt}}
\put(777.0,456.0){\rule[-0.200pt]{5.059pt}{0.400pt}}
\put(859,457.67){\rule{9.636pt}{0.400pt}}
\multiput(859.00,457.17)(20.000,1.000){2}{\rule{4.818pt}{0.400pt}}
\put(899,458.67){\rule{4.818pt}{0.400pt}}
\multiput(899.00,458.17)(10.000,1.000){2}{\rule{2.409pt}{0.400pt}}
\put(919,459.67){\rule{4.818pt}{0.400pt}}
\multiput(919.00,459.17)(10.000,1.000){2}{\rule{2.409pt}{0.400pt}}
\put(838.0,458.0){\rule[-0.200pt]{5.059pt}{0.400pt}}
\put(980,460.67){\rule{4.818pt}{0.400pt}}
\multiput(980.00,460.17)(10.000,1.000){2}{\rule{2.409pt}{0.400pt}}
\put(939.0,461.0){\rule[-0.200pt]{9.877pt}{0.400pt}}
\put(1021,461.67){\rule{4.818pt}{0.400pt}}
\multiput(1021.00,461.17)(10.000,1.000){2}{\rule{2.409pt}{0.400pt}}
\put(1000.0,462.0){\rule[-0.200pt]{5.059pt}{0.400pt}}
\put(1061,462.67){\rule{4.818pt}{0.400pt}}
\multiput(1061.00,462.17)(10.000,1.000){2}{\rule{2.409pt}{0.400pt}}
\put(1041.0,463.0){\rule[-0.200pt]{4.818pt}{0.400pt}}
\put(1102,463.67){\rule{9.636pt}{0.400pt}}
\multiput(1102.00,463.17)(20.000,1.000){2}{\rule{4.818pt}{0.400pt}}
\put(1081.0,464.0){\rule[-0.200pt]{5.059pt}{0.400pt}}
\put(1163,464.67){\rule{4.818pt}{0.400pt}}
\multiput(1163.00,464.17)(10.000,1.000){2}{\rule{2.409pt}{0.400pt}}
\put(1142.0,465.0){\rule[-0.200pt]{5.059pt}{0.400pt}}
\put(1243,465.67){\rule{5.059pt}{0.400pt}}
\multiput(1243.00,465.17)(10.500,1.000){2}{\rule{2.529pt}{0.400pt}}
\put(1183.0,466.0){\rule[-0.200pt]{14.454pt}{0.400pt}}
\multiput(1284.58,448.24)(0.496,-5.618){37}{\rule{0.119pt}{4.520pt}}
\multiput(1283.17,457.62)(20.000,-211.619){2}{\rule{0.400pt}{2.260pt}}
\put(1264.0,467.0){\rule[-0.200pt]{4.818pt}{0.400pt}}
\put(1304.0,113.0){\rule[-0.200pt]{0.400pt}{32.040pt}}
\put(1306,767){\makebox(0,0)[r]{"Neutralino LSP"}}
\multiput(1328,767)(20.756,0.000){4}{\usebox{\plotpoint}}
\put(1394,767){\usebox{\plotpoint}}
\put(238,689){\usebox{\plotpoint}}
\multiput(238,689)(11.027,17.584){4}{\usebox{\plotpoint}}
\multiput(275,748)(10.422,-17.949){3}{\usebox{\plotpoint}}
\multiput(311,686)(14.094,-15.237){3}{\usebox{\plotpoint}}
\multiput(348,646)(15.945,-13.287){2}{\usebox{\plotpoint}}
\multiput(384,616)(17.198,-11.620){2}{\usebox{\plotpoint}}
\multiput(421,591)(17.928,-10.458){2}{\usebox{\plotpoint}}
\multiput(457,570)(18.664,-9.080){2}{\usebox{\plotpoint}}
\multiput(494,552)(18.768,-8.863){2}{\usebox{\plotpoint}}
\multiput(530,535)(19.235,-7.798){2}{\usebox{\plotpoint}}
\multiput(567,520)(19.344,-7.523){2}{\usebox{\plotpoint}}
\multiput(603,506)(19.582,-6.880){2}{\usebox{\plotpoint}}
\multiput(640,493)(19.522,-7.049){2}{\usebox{\plotpoint}}
\multiput(676,480)(19.690,-6.563){2}{\usebox{\plotpoint}}
\put(729.92,462.67){\usebox{\plotpoint}}
\multiput(749,457)(19.522,-7.049){2}{\usebox{\plotpoint}}
\multiput(785,444)(19.743,-6.403){2}{\usebox{\plotpoint}}
\multiput(822,432)(19.690,-6.563){2}{\usebox{\plotpoint}}
\multiput(858,420)(19.582,-6.880){2}{\usebox{\plotpoint}}
\multiput(895,407)(19.690,-6.563){2}{\usebox{\plotpoint}}
\multiput(931,395)(19.412,-7.345){2}{\usebox{\plotpoint}}
\multiput(968,381)(18.967,-8.430){2}{\usebox{\plotpoint}}
\put(1022.35,356.57){\usebox{\plotpoint}}
\multiput(1041,348)(18.144,-10.080){2}{\usebox{\plotpoint}}
\multiput(1077,328)(17.198,-11.620){3}{\usebox{\plotpoint}}
\multiput(1114,303)(15.945,-13.287){2}{\usebox{\plotpoint}}
\multiput(1150,273)(13.009,-16.173){3}{\usebox{\plotpoint}}
\multiput(1187,227)(7.936,-19.178){4}{\usebox{\plotpoint}}
\multiput(1223,140)(0.000,-20.756){2}{\usebox{\plotpoint}}
\put(1223,113){\usebox{\plotpoint}}
\sbox{\plotpoint}{\rule[-0.400pt]{0.800pt}{0.800pt}}%
\put(1306,722){\makebox(0,0)[r]{"$\tilde\nu +$ LSP"}}
\put(1328.0,722.0){\rule[-0.400pt]{15.899pt}{0.800pt}}
\put(238,740){\usebox{\plotpoint}}
\multiput(239.41,740.00)(0.503,0.801){67}{\rule{0.121pt}{1.476pt}}
\multiput(236.34,740.00)(37.000,55.937){2}{\rule{0.800pt}{0.738pt}}
\multiput(276.41,792.63)(0.503,-0.838){65}{\rule{0.121pt}{1.533pt}}
\multiput(273.34,795.82)(36.000,-56.817){2}{\rule{0.800pt}{0.767pt}}
\multiput(312.41,734.67)(0.503,-0.525){67}{\rule{0.121pt}{1.043pt}}
\multiput(309.34,736.83)(37.000,-36.835){2}{\rule{0.800pt}{0.522pt}}
\multiput(348.00,698.09)(0.600,-0.503){53}{\rule{1.160pt}{0.121pt}}
\multiput(348.00,698.34)(33.592,-30.000){2}{\rule{0.580pt}{0.800pt}}
\multiput(384.00,668.09)(0.811,-0.505){39}{\rule{1.487pt}{0.122pt}}
\multiput(384.00,668.34)(33.914,-23.000){2}{\rule{0.743pt}{0.800pt}}
\multiput(421.00,645.09)(0.963,-0.506){31}{\rule{1.716pt}{0.122pt}}
\multiput(421.00,645.34)(32.439,-19.000){2}{\rule{0.858pt}{0.800pt}}
\multiput(457.00,626.09)(1.113,-0.507){27}{\rule{1.941pt}{0.122pt}}
\multiput(457.00,626.34)(32.971,-17.000){2}{\rule{0.971pt}{0.800pt}}
\multiput(494.00,609.09)(1.236,-0.508){23}{\rule{2.120pt}{0.122pt}}
\multiput(494.00,609.34)(31.600,-15.000){2}{\rule{1.060pt}{0.800pt}}
\multiput(530.00,594.08)(1.484,-0.509){19}{\rule{2.477pt}{0.123pt}}
\multiput(530.00,594.34)(31.859,-13.000){2}{\rule{1.238pt}{0.800pt}}
\multiput(567.00,581.08)(1.575,-0.511){17}{\rule{2.600pt}{0.123pt}}
\multiput(567.00,581.34)(30.604,-12.000){2}{\rule{1.300pt}{0.800pt}}
\multiput(603.00,569.08)(1.484,-0.509){19}{\rule{2.477pt}{0.123pt}}
\multiput(603.00,569.34)(31.859,-13.000){2}{\rule{1.238pt}{0.800pt}}
\multiput(640.00,556.08)(2.528,-0.520){9}{\rule{3.800pt}{0.125pt}}
\multiput(640.00,556.34)(28.113,-8.000){2}{\rule{1.900pt}{0.800pt}}
\multiput(676.00,548.08)(2.189,-0.516){11}{\rule{3.400pt}{0.124pt}}
\multiput(676.00,548.34)(28.943,-9.000){2}{\rule{1.700pt}{0.800pt}}
\multiput(712.00,539.08)(2.252,-0.516){11}{\rule{3.489pt}{0.124pt}}
\multiput(712.00,539.34)(29.759,-9.000){2}{\rule{1.744pt}{0.800pt}}
\multiput(749.00,530.08)(2.528,-0.520){9}{\rule{3.800pt}{0.125pt}}
\multiput(749.00,530.34)(28.113,-8.000){2}{\rule{1.900pt}{0.800pt}}
\multiput(785.00,522.08)(2.252,-0.516){11}{\rule{3.489pt}{0.124pt}}
\multiput(785.00,522.34)(29.759,-9.000){2}{\rule{1.744pt}{0.800pt}}
\multiput(822.00,513.08)(2.189,-0.516){11}{\rule{3.400pt}{0.124pt}}
\multiput(822.00,513.34)(28.943,-9.000){2}{\rule{1.700pt}{0.800pt}}
\multiput(858.00,504.07)(3.923,-0.536){5}{\rule{5.133pt}{0.129pt}}
\multiput(858.00,504.34)(26.346,-6.000){2}{\rule{2.567pt}{0.800pt}}
\multiput(895.00,498.08)(3.015,-0.526){7}{\rule{4.314pt}{0.127pt}}
\multiput(895.00,498.34)(27.045,-7.000){2}{\rule{2.157pt}{0.800pt}}
\multiput(931.00,491.08)(1.786,-0.512){15}{\rule{2.891pt}{0.123pt}}
\multiput(931.00,491.34)(31.000,-11.000){2}{\rule{1.445pt}{0.800pt}}
\put(968,478.34){\rule{7.400pt}{0.800pt}}
\multiput(968.00,480.34)(20.641,-4.000){2}{\rule{3.700pt}{0.800pt}}
\multiput(1004.00,476.08)(2.252,-0.516){11}{\rule{3.489pt}{0.124pt}}
\multiput(1004.00,476.34)(29.759,-9.000){2}{\rule{1.744pt}{0.800pt}}
\multiput(1041.00,467.08)(2.189,-0.516){11}{\rule{3.400pt}{0.124pt}}
\multiput(1041.00,467.34)(28.943,-9.000){2}{\rule{1.700pt}{0.800pt}}
\multiput(1077.00,458.08)(1.990,-0.514){13}{\rule{3.160pt}{0.124pt}}
\multiput(1077.00,458.34)(30.441,-10.000){2}{\rule{1.580pt}{0.800pt}}
\multiput(1114.00,448.08)(1.736,-0.512){15}{\rule{2.818pt}{0.123pt}}
\multiput(1114.00,448.34)(30.151,-11.000){2}{\rule{1.409pt}{0.800pt}}
\multiput(1150.00,437.09)(0.990,-0.506){31}{\rule{1.758pt}{0.122pt}}
\multiput(1150.00,437.34)(33.351,-19.000){2}{\rule{0.879pt}{0.800pt}}
\multiput(1187.00,418.09)(0.561,-0.503){57}{\rule{1.100pt}{0.121pt}}
\multiput(1187.00,418.34)(33.717,-32.000){2}{\rule{0.550pt}{0.800pt}}
\multiput(1223.00,386.09)(0.938,-0.505){33}{\rule{1.680pt}{0.122pt}}
\multiput(1223.00,386.34)(33.513,-20.000){2}{\rule{0.840pt}{0.800pt}}
\multiput(1261.41,363.48)(0.503,-0.554){65}{\rule{0.121pt}{1.089pt}}
\multiput(1258.34,365.74)(36.000,-37.740){2}{\rule{0.800pt}{0.544pt}}
\multiput(1297.41,314.78)(0.503,-1.891){67}{\rule{0.121pt}{3.184pt}}
\multiput(1294.34,321.39)(37.000,-131.392){2}{\rule{0.800pt}{1.592pt}}
\put(1333.0,113.0){\rule[-0.400pt]{0.800pt}{18.549pt}}
\sbox{\plotpoint}{\rule[-0.500pt]{1.000pt}{1.000pt}}%
\put(1306,677){\makebox(0,0)[r]{"Total VLSP"}}
\multiput(1328,677)(20.756,0.000){4}{\usebox{\plotpoint}}
\put(1394,677){\usebox{\plotpoint}}
\put(238,733){\usebox{\plotpoint}}
\multiput(238,733)(11.163,17.498){4}{\usebox{\plotpoint}}
\multiput(275,791)(10.679,-17.798){3}{\usebox{\plotpoint}}
\multiput(311,731)(14.094,-15.237){3}{\usebox{\plotpoint}}
\multiput(348,691)(16.163,-13.021){2}{\usebox{\plotpoint}}
\multiput(384,662)(17.413,-11.295){2}{\usebox{\plotpoint}}
\multiput(421,638)(18.144,-10.080){2}{\usebox{\plotpoint}}
\multiput(457,618)(19.051,-8.238){2}{\usebox{\plotpoint}}
\multiput(494,602)(18.967,-8.430){2}{\usebox{\plotpoint}}
\multiput(530,586)(19.235,-7.798){2}{\usebox{\plotpoint}}
\multiput(567,571)(19.522,-7.049){2}{\usebox{\plotpoint}}
\multiput(603,558)(20.037,-5.415){2}{\usebox{\plotpoint}}
\put(657.68,542.11){\usebox{\plotpoint}}
\multiput(676,536)(20.374,-3.962){2}{\usebox{\plotpoint}}
\multiput(712,529)(19.895,-5.915){2}{\usebox{\plotpoint}}
\multiput(749,518)(20.136,-5.034){2}{\usebox{\plotpoint}}
\multiput(785,509)(20.167,-4.906){2}{\usebox{\plotpoint}}
\put(838.29,495.02){\usebox{\plotpoint}}
\multiput(858,489)(20.394,-3.858){2}{\usebox{\plotpoint}}
\multiput(895,482)(20.136,-5.034){2}{\usebox{\plotpoint}}
\multiput(931,473)(20.167,-4.906){2}{\usebox{\plotpoint}}
\multiput(968,464)(20.136,-5.034){2}{\usebox{\plotpoint}}
\multiput(1004,455)(19.051,-8.238){2}{\usebox{\plotpoint}}
\multiput(1041,439)(19.522,-7.049){2}{\usebox{\plotpoint}}
\put(1096.50,419.15){\usebox{\plotpoint}}
\multiput(1114,413)(18.967,-8.430){2}{\usebox{\plotpoint}}
\multiput(1150,397)(18.463,-9.481){2}{\usebox{\plotpoint}}
\multiput(1187,378)(19.998,-5.555){2}{\usebox{\plotpoint}}
\multiput(1223,368)(17.198,-11.620){2}{\usebox{\plotpoint}}
\multiput(1260,343)(11.814,-17.065){3}{\usebox{\plotpoint}}
\multiput(1296,291)(7.140,-19.489){5}{\usebox{\plotpoint}}
\multiput(1333,190)(0.000,-20.756){4}{\usebox{\plotpoint}}
\put(1333,113){\usebox{\plotpoint}}
\end{picture}

\vskip 2 true cm

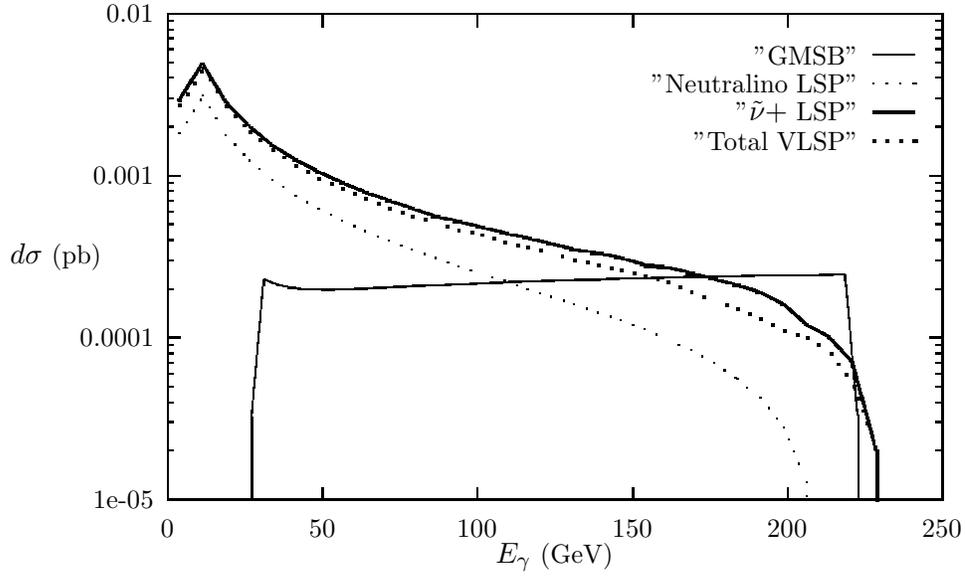
\centerline {\bf Fig. 5}

\noindent Energy distribution of the emitted photon with angular cuts as stated
in the text. The parameter space for GMSB is the same as in Figure 1, 
while for MSSM, the three curves correspond to the parameter sets 
shown in Figure 4.

\newpage
\setlength{\unitlength}{0.240900pt}
\ifx\plotpoint\undefined\newsavebox{\plotpoint}\fi
\sbox{\plotpoint}{\rule[-0.200pt]{0.400pt}{0.400pt}}%
\begin{picture}(1500,900)(0,0)
\font\gnuplot=cmr10 at 10pt
\gnuplot
\sbox{\plotpoint}{\rule[-0.200pt]{0.400pt}{0.400pt}}%
\put(220.0,113.0){\rule[-0.200pt]{0.400pt}{184.048pt}}
\put(220.0,113.0){\rule[-0.200pt]{4.818pt}{0.400pt}}
\put(198,113){\makebox(0,0)[r]{1e-06}}
\put(1416.0,113.0){\rule[-0.200pt]{4.818pt}{0.400pt}}
\put(220.0,170.0){\rule[-0.200pt]{2.409pt}{0.400pt}}
\put(1426.0,170.0){\rule[-0.200pt]{2.409pt}{0.400pt}}
\put(220.0,204.0){\rule[-0.200pt]{2.409pt}{0.400pt}}
\put(1426.0,204.0){\rule[-0.200pt]{2.409pt}{0.400pt}}
\put(220.0,228.0){\rule[-0.200pt]{2.409pt}{0.400pt}}
\put(1426.0,228.0){\rule[-0.200pt]{2.409pt}{0.400pt}}
\put(220.0,247.0){\rule[-0.200pt]{2.409pt}{0.400pt}}
\put(1426.0,247.0){\rule[-0.200pt]{2.409pt}{0.400pt}}
\put(220.0,262.0){\rule[-0.200pt]{2.409pt}{0.400pt}}
\put(1426.0,262.0){\rule[-0.200pt]{2.409pt}{0.400pt}}
\put(220.0,274.0){\rule[-0.200pt]{2.409pt}{0.400pt}}
\put(1426.0,274.0){\rule[-0.200pt]{2.409pt}{0.400pt}}
\put(220.0,285.0){\rule[-0.200pt]{2.409pt}{0.400pt}}
\put(1426.0,285.0){\rule[-0.200pt]{2.409pt}{0.400pt}}
\put(220.0,295.0){\rule[-0.200pt]{2.409pt}{0.400pt}}
\put(1426.0,295.0){\rule[-0.200pt]{2.409pt}{0.400pt}}
\put(220.0,304.0){\rule[-0.200pt]{4.818pt}{0.400pt}}
\put(198,304){\makebox(0,0)[r]{1e-05}}
\put(1416.0,304.0){\rule[-0.200pt]{4.818pt}{0.400pt}}
\put(220.0,361.0){\rule[-0.200pt]{2.409pt}{0.400pt}}
\put(1426.0,361.0){\rule[-0.200pt]{2.409pt}{0.400pt}}
\put(220.0,395.0){\rule[-0.200pt]{2.409pt}{0.400pt}}
\put(1426.0,395.0){\rule[-0.200pt]{2.409pt}{0.400pt}}
\put(220.0,419.0){\rule[-0.200pt]{2.409pt}{0.400pt}}
\put(1426.0,419.0){\rule[-0.200pt]{2.409pt}{0.400pt}}
\put(220.0,438.0){\rule[-0.200pt]{2.409pt}{0.400pt}}
\put(1426.0,438.0){\rule[-0.200pt]{2.409pt}{0.400pt}}
\put(220.0,453.0){\rule[-0.200pt]{2.409pt}{0.400pt}}
\put(1426.0,453.0){\rule[-0.200pt]{2.409pt}{0.400pt}}
\put(220.0,465.0){\rule[-0.200pt]{2.409pt}{0.400pt}}
\put(1426.0,465.0){\rule[-0.200pt]{2.409pt}{0.400pt}}
\put(220.0,476.0){\rule[-0.200pt]{2.409pt}{0.400pt}}
\put(1426.0,476.0){\rule[-0.200pt]{2.409pt}{0.400pt}}
\put(220.0,486.0){\rule[-0.200pt]{2.409pt}{0.400pt}}
\put(1426.0,486.0){\rule[-0.200pt]{2.409pt}{0.400pt}}
\put(220.0,495.0){\rule[-0.200pt]{4.818pt}{0.400pt}}
\put(198,495){\makebox(0,0)[r]{0.0001}}
\put(1416.0,495.0){\rule[-0.200pt]{4.818pt}{0.400pt}}
\put(220.0,552.0){\rule[-0.200pt]{2.409pt}{0.400pt}}
\put(1426.0,552.0){\rule[-0.200pt]{2.409pt}{0.400pt}}
\put(220.0,586.0){\rule[-0.200pt]{2.409pt}{0.400pt}}
\put(1426.0,586.0){\rule[-0.200pt]{2.409pt}{0.400pt}}
\put(220.0,610.0){\rule[-0.200pt]{2.409pt}{0.400pt}}
\put(1426.0,610.0){\rule[-0.200pt]{2.409pt}{0.400pt}}
\put(220.0,629.0){\rule[-0.200pt]{2.409pt}{0.400pt}}
\put(1426.0,629.0){\rule[-0.200pt]{2.409pt}{0.400pt}}
\put(220.0,644.0){\rule[-0.200pt]{2.409pt}{0.400pt}}
\put(1426.0,644.0){\rule[-0.200pt]{2.409pt}{0.400pt}}
\put(220.0,656.0){\rule[-0.200pt]{2.409pt}{0.400pt}}
\put(1426.0,656.0){\rule[-0.200pt]{2.409pt}{0.400pt}}
\put(220.0,667.0){\rule[-0.200pt]{2.409pt}{0.400pt}}
\put(1426.0,667.0){\rule[-0.200pt]{2.409pt}{0.400pt}}
\put(220.0,677.0){\rule[-0.200pt]{2.409pt}{0.400pt}}
\put(1426.0,677.0){\rule[-0.200pt]{2.409pt}{0.400pt}}
\put(220.0,686.0){\rule[-0.200pt]{4.818pt}{0.400pt}}
\put(198,686){\makebox(0,0)[r]{0.001}}
\put(1416.0,686.0){\rule[-0.200pt]{4.818pt}{0.400pt}}
\put(220.0,743.0){\rule[-0.200pt]{2.409pt}{0.400pt}}
\put(1426.0,743.0){\rule[-0.200pt]{2.409pt}{0.400pt}}
\put(220.0,777.0){\rule[-0.200pt]{2.409pt}{0.400pt}}
\put(1426.0,777.0){\rule[-0.200pt]{2.409pt}{0.400pt}}
\put(220.0,801.0){\rule[-0.200pt]{2.409pt}{0.400pt}}
\put(1426.0,801.0){\rule[-0.200pt]{2.409pt}{0.400pt}}
\put(220.0,820.0){\rule[-0.200pt]{2.409pt}{0.400pt}}
\put(1426.0,820.0){\rule[-0.200pt]{2.409pt}{0.400pt}}
\put(220.0,835.0){\rule[-0.200pt]{2.409pt}{0.400pt}}
\put(1426.0,835.0){\rule[-0.200pt]{2.409pt}{0.400pt}}
\put(220.0,847.0){\rule[-0.200pt]{2.409pt}{0.400pt}}
\put(1426.0,847.0){\rule[-0.200pt]{2.409pt}{0.400pt}}
\put(220.0,858.0){\rule[-0.200pt]{2.409pt}{0.400pt}}
\put(1426.0,858.0){\rule[-0.200pt]{2.409pt}{0.400pt}}
\put(220.0,868.0){\rule[-0.200pt]{2.409pt}{0.400pt}}
\put(1426.0,868.0){\rule[-0.200pt]{2.409pt}{0.400pt}}
\put(220.0,877.0){\rule[-0.200pt]{4.818pt}{0.400pt}}
\put(198,877){\makebox(0,0)[r]{0.01}}
\put(1416.0,877.0){\rule[-0.200pt]{4.818pt}{0.400pt}}
\put(220.0,113.0){\rule[-0.200pt]{0.400pt}{4.818pt}}
\put(220,68){\makebox(0,0){0}}
\put(220.0,857.0){\rule[-0.200pt]{0.400pt}{4.818pt}}
\put(355.0,113.0){\rule[-0.200pt]{0.400pt}{4.818pt}}
\put(355,68){\makebox(0,0){20}}
\put(355.0,857.0){\rule[-0.200pt]{0.400pt}{4.818pt}}
\put(490.0,113.0){\rule[-0.200pt]{0.400pt}{4.818pt}}
\put(490,68){\makebox(0,0){40}}
\put(490.0,857.0){\rule[-0.200pt]{0.400pt}{4.818pt}}
\put(625.0,113.0){\rule[-0.200pt]{0.400pt}{4.818pt}}
\put(625,68){\makebox(0,0){60}}
\put(625.0,857.0){\rule[-0.200pt]{0.400pt}{4.818pt}}
\put(760.0,113.0){\rule[-0.200pt]{0.400pt}{4.818pt}}
\put(760,68){\makebox(0,0){80}}
\put(760.0,857.0){\rule[-0.200pt]{0.400pt}{4.818pt}}
\put(896.0,113.0){\rule[-0.200pt]{0.400pt}{4.818pt}}
\put(896,68){\makebox(0,0){100}}
\put(896.0,857.0){\rule[-0.200pt]{0.400pt}{4.818pt}}
\put(1031.0,113.0){\rule[-0.200pt]{0.400pt}{4.818pt}}
\put(1031,68){\makebox(0,0){120}}
\put(1031.0,857.0){\rule[-0.200pt]{0.400pt}{4.818pt}}
\put(1166.0,113.0){\rule[-0.200pt]{0.400pt}{4.818pt}}
\put(1166,68){\makebox(0,0){140}}
\put(1166.0,857.0){\rule[-0.200pt]{0.400pt}{4.818pt}}
\put(1301.0,113.0){\rule[-0.200pt]{0.400pt}{4.818pt}}
\put(1301,68){\makebox(0,0){160}}
\put(1301.0,857.0){\rule[-0.200pt]{0.400pt}{4.818pt}}
\put(1436.0,113.0){\rule[-0.200pt]{0.400pt}{4.818pt}}
\put(1436,68){\makebox(0,0){180}}
\put(1436.0,857.0){\rule[-0.200pt]{0.400pt}{4.818pt}}
\put(220.0,113.0){\rule[-0.200pt]{292.934pt}{0.400pt}}
\put(1436.0,113.0){\rule[-0.200pt]{0.400pt}{184.048pt}}
\put(220.0,877.0){\rule[-0.200pt]{292.934pt}{0.400pt}}
\put(15,495){\makebox(0,0){$d\sigma$ (pb)}}
\put(828,23){\makebox(0,0){$\theta_{\gamma}$ (deg)}}
\put(220.0,113.0){\rule[-0.200pt]{0.400pt}{184.048pt}}
\put(1306,812){\makebox(0,0)[r]{"GMSB"}}
\put(1328.0,812.0){\rule[-0.200pt]{15.899pt}{0.400pt}}
\put(230,204){\usebox{\plotpoint}}
\multiput(230.58,204.00)(0.496,2.281){37}{\rule{0.119pt}{1.900pt}}
\multiput(229.17,204.00)(20.000,86.056){2}{\rule{0.400pt}{0.950pt}}
\multiput(250.58,294.00)(0.496,1.177){39}{\rule{0.119pt}{1.033pt}}
\multiput(249.17,294.00)(21.000,46.855){2}{\rule{0.400pt}{0.517pt}}
\multiput(271.58,343.00)(0.496,0.651){37}{\rule{0.119pt}{0.620pt}}
\multiput(270.17,343.00)(20.000,24.713){2}{\rule{0.400pt}{0.310pt}}
\multiput(291.58,369.00)(0.496,0.676){37}{\rule{0.119pt}{0.640pt}}
\multiput(290.17,369.00)(20.000,25.672){2}{\rule{0.400pt}{0.320pt}}
\multiput(311.00,396.58)(0.588,0.495){31}{\rule{0.571pt}{0.119pt}}
\multiput(311.00,395.17)(18.816,17.000){2}{\rule{0.285pt}{0.400pt}}
\multiput(331.00,413.58)(0.657,0.494){29}{\rule{0.625pt}{0.119pt}}
\multiput(331.00,412.17)(19.703,16.000){2}{\rule{0.313pt}{0.400pt}}
\multiput(352.00,429.58)(0.588,0.495){31}{\rule{0.571pt}{0.119pt}}
\multiput(352.00,428.17)(18.816,17.000){2}{\rule{0.285pt}{0.400pt}}
\multiput(372.00,446.58)(0.841,0.492){21}{\rule{0.767pt}{0.119pt}}
\multiput(372.00,445.17)(18.409,12.000){2}{\rule{0.383pt}{0.400pt}}
\multiput(392.00,458.58)(0.814,0.493){23}{\rule{0.746pt}{0.119pt}}
\multiput(392.00,457.17)(19.451,13.000){2}{\rule{0.373pt}{0.400pt}}
\multiput(413.00,471.58)(0.920,0.492){19}{\rule{0.827pt}{0.118pt}}
\multiput(413.00,470.17)(18.283,11.000){2}{\rule{0.414pt}{0.400pt}}
\multiput(433.00,482.58)(0.920,0.492){19}{\rule{0.827pt}{0.118pt}}
\multiput(433.00,481.17)(18.283,11.000){2}{\rule{0.414pt}{0.400pt}}
\multiput(453.00,493.59)(1.286,0.488){13}{\rule{1.100pt}{0.117pt}}
\multiput(453.00,492.17)(17.717,8.000){2}{\rule{0.550pt}{0.400pt}}
\multiput(473.00,501.58)(1.069,0.491){17}{\rule{0.940pt}{0.118pt}}
\multiput(473.00,500.17)(19.049,10.000){2}{\rule{0.470pt}{0.400pt}}
\multiput(494.00,511.59)(1.286,0.488){13}{\rule{1.100pt}{0.117pt}}
\multiput(494.00,510.17)(17.717,8.000){2}{\rule{0.550pt}{0.400pt}}
\multiput(514.00,519.59)(1.286,0.488){13}{\rule{1.100pt}{0.117pt}}
\multiput(514.00,518.17)(17.717,8.000){2}{\rule{0.550pt}{0.400pt}}
\multiput(534.00,527.59)(1.756,0.482){9}{\rule{1.433pt}{0.116pt}}
\multiput(534.00,526.17)(17.025,6.000){2}{\rule{0.717pt}{0.400pt}}
\multiput(554.00,533.59)(1.352,0.488){13}{\rule{1.150pt}{0.117pt}}
\multiput(554.00,532.17)(18.613,8.000){2}{\rule{0.575pt}{0.400pt}}
\multiput(575.00,541.59)(2.157,0.477){7}{\rule{1.700pt}{0.115pt}}
\multiput(575.00,540.17)(16.472,5.000){2}{\rule{0.850pt}{0.400pt}}
\multiput(595.00,546.59)(2.157,0.477){7}{\rule{1.700pt}{0.115pt}}
\multiput(595.00,545.17)(16.472,5.000){2}{\rule{0.850pt}{0.400pt}}
\multiput(615.00,551.58)(2.400,0.493){23}{\rule{1.977pt}{0.119pt}}
\multiput(615.00,550.17)(56.897,13.000){2}{\rule{0.988pt}{0.400pt}}
\multiput(676.00,564.61)(4.258,0.447){3}{\rule{2.767pt}{0.108pt}}
\multiput(676.00,563.17)(14.258,3.000){2}{\rule{1.383pt}{0.400pt}}
\multiput(696.00,567.60)(2.967,0.468){5}{\rule{2.200pt}{0.113pt}}
\multiput(696.00,566.17)(16.434,4.000){2}{\rule{1.100pt}{0.400pt}}
\put(717,570.67){\rule{4.818pt}{0.400pt}}
\multiput(717.00,570.17)(10.000,1.000){2}{\rule{2.409pt}{0.400pt}}
\put(737,572.17){\rule{4.100pt}{0.400pt}}
\multiput(737.00,571.17)(11.490,2.000){2}{\rule{2.050pt}{0.400pt}}
\put(757,573.67){\rule{4.818pt}{0.400pt}}
\multiput(757.00,573.17)(10.000,1.000){2}{\rule{2.409pt}{0.400pt}}
\put(777,574.67){\rule{5.059pt}{0.400pt}}
\multiput(777.00,574.17)(10.500,1.000){2}{\rule{2.529pt}{0.400pt}}
\put(798,575.67){\rule{4.818pt}{0.400pt}}
\multiput(798.00,575.17)(10.000,1.000){2}{\rule{2.409pt}{0.400pt}}
\put(838,575.67){\rule{4.818pt}{0.400pt}}
\multiput(838.00,576.17)(10.000,-1.000){2}{\rule{2.409pt}{0.400pt}}
\put(858,574.67){\rule{5.059pt}{0.400pt}}
\multiput(858.00,575.17)(10.500,-1.000){2}{\rule{2.529pt}{0.400pt}}
\put(879,573.67){\rule{4.818pt}{0.400pt}}
\multiput(879.00,574.17)(10.000,-1.000){2}{\rule{2.409pt}{0.400pt}}
\put(899,572.17){\rule{4.100pt}{0.400pt}}
\multiput(899.00,573.17)(11.490,-2.000){2}{\rule{2.050pt}{0.400pt}}
\put(919,570.17){\rule{4.100pt}{0.400pt}}
\multiput(919.00,571.17)(11.490,-2.000){2}{\rule{2.050pt}{0.400pt}}
\multiput(939.00,568.95)(4.481,-0.447){3}{\rule{2.900pt}{0.108pt}}
\multiput(939.00,569.17)(14.981,-3.000){2}{\rule{1.450pt}{0.400pt}}
\multiput(960.00,565.95)(4.258,-0.447){3}{\rule{2.767pt}{0.108pt}}
\multiput(960.00,566.17)(14.258,-3.000){2}{\rule{1.383pt}{0.400pt}}
\multiput(980.00,562.95)(4.258,-0.447){3}{\rule{2.767pt}{0.108pt}}
\multiput(980.00,563.17)(14.258,-3.000){2}{\rule{1.383pt}{0.400pt}}
\multiput(1000.00,559.93)(2.269,-0.477){7}{\rule{1.780pt}{0.115pt}}
\multiput(1000.00,560.17)(17.306,-5.000){2}{\rule{0.890pt}{0.400pt}}
\multiput(1021.00,554.93)(2.157,-0.477){7}{\rule{1.700pt}{0.115pt}}
\multiput(1021.00,555.17)(16.472,-5.000){2}{\rule{0.850pt}{0.400pt}}
\multiput(1041.00,549.93)(2.157,-0.477){7}{\rule{1.700pt}{0.115pt}}
\multiput(1041.00,550.17)(16.472,-5.000){2}{\rule{0.850pt}{0.400pt}}
\multiput(1061.00,544.93)(2.157,-0.477){7}{\rule{1.700pt}{0.115pt}}
\multiput(1061.00,545.17)(16.472,-5.000){2}{\rule{0.850pt}{0.400pt}}
\multiput(1081.00,539.93)(1.352,-0.488){13}{\rule{1.150pt}{0.117pt}}
\multiput(1081.00,540.17)(18.613,-8.000){2}{\rule{0.575pt}{0.400pt}}
\multiput(1102.00,531.93)(2.157,-0.477){7}{\rule{1.700pt}{0.115pt}}
\multiput(1102.00,532.17)(16.472,-5.000){2}{\rule{0.850pt}{0.400pt}}
\multiput(1122.00,526.93)(1.135,-0.489){15}{\rule{0.989pt}{0.118pt}}
\multiput(1122.00,527.17)(17.948,-9.000){2}{\rule{0.494pt}{0.400pt}}
\multiput(1142.00,517.93)(1.286,-0.488){13}{\rule{1.100pt}{0.117pt}}
\multiput(1142.00,518.17)(17.717,-8.000){2}{\rule{0.550pt}{0.400pt}}
\multiput(1162.00,509.93)(1.194,-0.489){15}{\rule{1.033pt}{0.118pt}}
\multiput(1162.00,510.17)(18.855,-9.000){2}{\rule{0.517pt}{0.400pt}}
\multiput(1183.00,500.93)(1.135,-0.489){15}{\rule{0.989pt}{0.118pt}}
\multiput(1183.00,501.17)(17.948,-9.000){2}{\rule{0.494pt}{0.400pt}}
\multiput(1203.00,491.92)(1.017,-0.491){17}{\rule{0.900pt}{0.118pt}}
\multiput(1203.00,492.17)(18.132,-10.000){2}{\rule{0.450pt}{0.400pt}}
\multiput(1223.00,481.92)(0.841,-0.492){21}{\rule{0.767pt}{0.119pt}}
\multiput(1223.00,482.17)(18.409,-12.000){2}{\rule{0.383pt}{0.400pt}}
\multiput(1243.00,469.92)(0.967,-0.492){19}{\rule{0.864pt}{0.118pt}}
\multiput(1243.00,470.17)(19.207,-11.000){2}{\rule{0.432pt}{0.400pt}}
\multiput(1264.00,458.92)(0.625,-0.494){29}{\rule{0.600pt}{0.119pt}}
\multiput(1264.00,459.17)(18.755,-16.000){2}{\rule{0.300pt}{0.400pt}}
\multiput(1284.00,442.92)(0.774,-0.493){23}{\rule{0.715pt}{0.119pt}}
\multiput(1284.00,443.17)(18.515,-13.000){2}{\rule{0.358pt}{0.400pt}}
\multiput(1304.00,429.92)(0.583,-0.495){33}{\rule{0.567pt}{0.119pt}}
\multiput(1304.00,430.17)(19.824,-18.000){2}{\rule{0.283pt}{0.400pt}}
\multiput(1325.00,411.92)(0.525,-0.495){35}{\rule{0.521pt}{0.119pt}}
\multiput(1325.00,412.17)(18.919,-19.000){2}{\rule{0.261pt}{0.400pt}}
\multiput(1345.58,391.76)(0.496,-0.549){37}{\rule{0.119pt}{0.540pt}}
\multiput(1344.17,392.88)(20.000,-20.879){2}{\rule{0.400pt}{0.270pt}}
\multiput(1365.58,369.18)(0.496,-0.727){37}{\rule{0.119pt}{0.680pt}}
\multiput(1364.17,370.59)(20.000,-27.589){2}{\rule{0.400pt}{0.340pt}}
\multiput(1385.58,339.03)(0.496,-1.080){39}{\rule{0.119pt}{0.957pt}}
\multiput(1384.17,341.01)(21.000,-43.013){2}{\rule{0.400pt}{0.479pt}}
\multiput(1406.58,290.20)(0.496,-2.255){37}{\rule{0.119pt}{1.880pt}}
\multiput(1405.17,294.10)(20.000,-85.098){2}{\rule{0.400pt}{0.940pt}}
\put(818.0,577.0){\rule[-0.200pt]{4.818pt}{0.400pt}}
\put(1306,767){\makebox(0,0)[r]{"Neutralino LSP"}}
\multiput(1328,767)(20.756,0.000){4}{\usebox{\plotpoint}}
\put(1394,767){\usebox{\plotpoint}}
\put(266,642){\usebox{\plotpoint}}
\multiput(266,642)(14.434,-14.915){3}{\usebox{\plotpoint}}
\multiput(296,611)(15.427,-13.885){2}{\usebox{\plotpoint}}
\put(343.04,573.01){\usebox{\plotpoint}}
\multiput(357,564)(18.058,-10.233){2}{\usebox{\plotpoint}}
\multiput(387,547)(18.916,-8.543){2}{\usebox{\plotpoint}}
\put(435.33,526.07){\usebox{\plotpoint}}
\multiput(448,521)(20.055,-5.348){2}{\usebox{\plotpoint}}
\put(494.88,508.10){\usebox{\plotpoint}}
\multiput(509,504)(20.213,-4.716){2}{\usebox{\plotpoint}}
\put(555.34,493.31){\usebox{\plotpoint}}
\multiput(570,490)(20.473,-3.412){2}{\usebox{\plotpoint}}
\put(616.50,481.70){\usebox{\plotpoint}}
\multiput(630,479)(20.585,-2.656){2}{\usebox{\plotpoint}}
\put(678.23,474.43){\usebox{\plotpoint}}
\multiput(691,474)(20.659,-1.999){2}{\usebox{\plotpoint}}
\put(740.19,468.58){\usebox{\plotpoint}}
\multiput(752,467)(20.744,-0.691){2}{\usebox{\plotpoint}}
\put(802.29,467.31){\usebox{\plotpoint}}
\put(823.00,467.33){\usebox{\plotpoint}}
\multiput(843,466)(20.756,0.000){2}{\usebox{\plotpoint}}
\put(885.20,466.75){\usebox{\plotpoint}}
\multiput(904,468)(20.710,1.381){2}{\usebox{\plotpoint}}
\put(947.33,470.86){\usebox{\plotpoint}}
\multiput(965,472)(20.573,2.743){2}{\usebox{\plotpoint}}
\put(1009.17,477.83){\usebox{\plotpoint}}
\multiput(1026,480)(20.573,2.743){2}{\usebox{\plotpoint}}
\put(1070.64,487.42){\usebox{\plotpoint}}
\multiput(1086,491)(20.377,3.944){2}{\usebox{\plotpoint}}
\put(1131.63,499.93){\usebox{\plotpoint}}
\multiput(1147,503)(19.753,6.372){2}{\usebox{\plotpoint}}
\put(1191.43,517.03){\usebox{\plotpoint}}
\multiput(1208,522)(19.487,7.145){2}{\usebox{\plotpoint}}
\put(1250.14,537.70){\usebox{\plotpoint}}
\multiput(1269,545)(17.798,10.679){2}{\usebox{\plotpoint}}
\multiput(1299,563)(17.441,11.252){2}{\usebox{\plotpoint}}
\multiput(1330,583)(15.173,14.162){2}{\usebox{\plotpoint}}
\multiput(1360,611)(14.196,15.142){2}{\usebox{\plotpoint}}
\put(1390,643){\usebox{\plotpoint}}
\sbox{\plotpoint}{\rule[-0.400pt]{0.800pt}{0.800pt}}%
\put(1306,722){\makebox(0,0)[r]{"$\tilde\nu +$ LSP"}}
\put(1328.0,722.0){\rule[-0.400pt]{15.899pt}{0.800pt}}
\put(266,721){\usebox{\plotpoint}}
\multiput(266.00,719.09)(0.534,-0.504){49}{\rule{1.057pt}{0.121pt}}
\multiput(266.00,719.34)(27.806,-28.000){2}{\rule{0.529pt}{0.800pt}}
\multiput(296.00,691.09)(0.497,-0.503){53}{\rule{1.000pt}{0.121pt}}
\multiput(296.00,691.34)(27.924,-30.000){2}{\rule{0.500pt}{0.800pt}}
\multiput(326.00,661.09)(0.743,-0.505){35}{\rule{1.381pt}{0.122pt}}
\multiput(326.00,661.34)(28.134,-21.000){2}{\rule{0.690pt}{0.800pt}}
\multiput(357.00,640.09)(1.024,-0.508){23}{\rule{1.800pt}{0.122pt}}
\multiput(357.00,640.34)(26.264,-15.000){2}{\rule{0.900pt}{0.800pt}}
\multiput(387.00,625.09)(1.059,-0.508){23}{\rule{1.853pt}{0.122pt}}
\multiput(387.00,625.34)(27.153,-15.000){2}{\rule{0.927pt}{0.800pt}}
\multiput(418.00,610.08)(1.437,-0.512){15}{\rule{2.382pt}{0.123pt}}
\multiput(418.00,610.34)(25.056,-11.000){2}{\rule{1.191pt}{0.800pt}}
\multiput(448.00,599.08)(1.601,-0.514){13}{\rule{2.600pt}{0.124pt}}
\multiput(448.00,599.34)(24.604,-10.000){2}{\rule{1.300pt}{0.800pt}}
\multiput(478.00,589.08)(2.163,-0.520){9}{\rule{3.300pt}{0.125pt}}
\multiput(478.00,589.34)(24.151,-8.000){2}{\rule{1.650pt}{0.800pt}}
\multiput(509.00,581.08)(1.810,-0.516){11}{\rule{2.867pt}{0.124pt}}
\multiput(509.00,581.34)(24.050,-9.000){2}{\rule{1.433pt}{0.800pt}}
\multiput(539.00,572.07)(3.253,-0.536){5}{\rule{4.333pt}{0.129pt}}
\multiput(539.00,572.34)(22.006,-6.000){2}{\rule{2.167pt}{0.800pt}}
\put(570,564.34){\rule{6.200pt}{0.800pt}}
\multiput(570.00,566.34)(17.132,-4.000){2}{\rule{3.100pt}{0.800pt}}
\multiput(600.00,562.08)(2.490,-0.526){7}{\rule{3.629pt}{0.127pt}}
\multiput(600.00,562.34)(22.469,-7.000){2}{\rule{1.814pt}{0.800pt}}
\multiput(630.00,555.06)(4.790,-0.560){3}{\rule{5.160pt}{0.135pt}}
\multiput(630.00,555.34)(20.290,-5.000){2}{\rule{2.580pt}{0.800pt}}
\put(691,548.34){\rule{6.400pt}{0.800pt}}
\multiput(691.00,550.34)(17.716,-4.000){2}{\rule{3.200pt}{0.800pt}}
\put(722,544.34){\rule{6.200pt}{0.800pt}}
\multiput(722.00,546.34)(17.132,-4.000){2}{\rule{3.100pt}{0.800pt}}
\put(661.0,552.0){\rule[-0.400pt]{7.227pt}{0.800pt}}
\multiput(813.00,542.06)(4.622,-0.560){3}{\rule{5.000pt}{0.135pt}}
\multiput(813.00,542.34)(19.622,-5.000){2}{\rule{2.500pt}{0.800pt}}
\multiput(843.00,540.38)(4.790,0.560){3}{\rule{5.160pt}{0.135pt}}
\multiput(843.00,537.34)(20.290,5.000){2}{\rule{2.580pt}{0.800pt}}
\put(752.0,544.0){\rule[-0.400pt]{14.695pt}{0.800pt}}
\put(934,544.34){\rule{6.400pt}{0.800pt}}
\multiput(934.00,542.34)(17.716,4.000){2}{\rule{3.200pt}{0.800pt}}
\put(965,548.34){\rule{6.200pt}{0.800pt}}
\multiput(965.00,546.34)(17.132,4.000){2}{\rule{3.100pt}{0.800pt}}
\multiput(995.00,553.38)(4.790,0.560){3}{\rule{5.160pt}{0.135pt}}
\multiput(995.00,550.34)(20.290,5.000){2}{\rule{2.580pt}{0.800pt}}
\put(1026,556.84){\rule{7.227pt}{0.800pt}}
\multiput(1026.00,555.34)(15.000,3.000){2}{\rule{3.613pt}{0.800pt}}
\multiput(1056.00,561.40)(2.090,0.520){9}{\rule{3.200pt}{0.125pt}}
\multiput(1056.00,558.34)(23.358,8.000){2}{\rule{1.600pt}{0.800pt}}
\multiput(1086.00,569.39)(3.253,0.536){5}{\rule{4.333pt}{0.129pt}}
\multiput(1086.00,566.34)(22.006,6.000){2}{\rule{2.167pt}{0.800pt}}
\put(1117,573.84){\rule{7.227pt}{0.800pt}}
\multiput(1117.00,572.34)(15.000,3.000){2}{\rule{3.613pt}{0.800pt}}
\multiput(1147.00,578.41)(1.349,0.511){17}{\rule{2.267pt}{0.123pt}}
\multiput(1147.00,575.34)(26.295,12.000){2}{\rule{1.133pt}{0.800pt}}
\multiput(1178.00,590.40)(1.601,0.514){13}{\rule{2.600pt}{0.124pt}}
\multiput(1178.00,587.34)(24.604,10.000){2}{\rule{1.300pt}{0.800pt}}
\multiput(1208.00,600.40)(1.810,0.516){11}{\rule{2.867pt}{0.124pt}}
\multiput(1208.00,597.34)(24.050,9.000){2}{\rule{1.433pt}{0.800pt}}
\multiput(1238.00,609.41)(1.059,0.508){23}{\rule{1.853pt}{0.122pt}}
\multiput(1238.00,606.34)(27.153,15.000){2}{\rule{0.927pt}{0.800pt}}
\multiput(1269.00,624.41)(1.024,0.508){23}{\rule{1.800pt}{0.122pt}}
\multiput(1269.00,621.34)(26.264,15.000){2}{\rule{0.900pt}{0.800pt}}
\multiput(1299.00,639.41)(0.708,0.505){37}{\rule{1.327pt}{0.122pt}}
\multiput(1299.00,636.34)(28.245,22.000){2}{\rule{0.664pt}{0.800pt}}
\multiput(1330.00,661.41)(0.534,0.504){49}{\rule{1.057pt}{0.121pt}}
\multiput(1330.00,658.34)(27.806,28.000){2}{\rule{0.529pt}{0.800pt}}
\multiput(1361.41,688.00)(0.503,0.514){53}{\rule{0.121pt}{1.027pt}}
\multiput(1358.34,688.00)(30.000,28.869){2}{\rule{0.800pt}{0.513pt}}
\put(874.0,544.0){\rule[-0.400pt]{14.454pt}{0.800pt}}
\sbox{\plotpoint}{\rule[-0.500pt]{1.000pt}{1.000pt}}%
\put(1306,677){\makebox(0,0)[r]{"Total VLSP"}}
\multiput(1328,677)(20.756,0.000){4}{\usebox{\plotpoint}}
\put(1394,677){\usebox{\plotpoint}}
\put(266,706){\usebox{\plotpoint}}
\multiput(266,706)(14.923,-14.426){3}{\usebox{\plotpoint}}
\multiput(296,677)(14.923,-14.426){2}{\usebox{\plotpoint}}
\put(342.32,635.89){\usebox{\plotpoint}}
\multiput(357,625)(18.564,-9.282){2}{\usebox{\plotpoint}}
\multiput(387,610)(19.141,-8.027){2}{\usebox{\plotpoint}}
\put(435.23,590.68){\usebox{\plotpoint}}
\multiput(448,586)(19.271,-7.708){2}{\usebox{\plotpoint}}
\put(493.56,568.98){\usebox{\plotpoint}}
\multiput(509,564)(20.573,-2.743){2}{\usebox{\plotpoint}}
\put(554.28,556.06){\usebox{\plotpoint}}
\multiput(570,552)(20.573,-2.743){2}{\usebox{\plotpoint}}
\put(615.10,543.47){\usebox{\plotpoint}}
\multiput(630,539)(20.756,0.000){2}{\usebox{\plotpoint}}
\put(676.50,536.42){\usebox{\plotpoint}}
\multiput(691,534)(20.756,0.000){2}{\usebox{\plotpoint}}
\put(738.34,531.28){\usebox{\plotpoint}}
\multiput(752,529)(20.756,0.000){2}{\usebox{\plotpoint}}
\put(800.41,529.00){\usebox{\plotpoint}}
\multiput(813,529)(20.756,0.000){2}{\usebox{\plotpoint}}
\put(862.68,529.00){\usebox{\plotpoint}}
\put(883.44,529.00){\usebox{\plotpoint}}
\multiput(904,529)(20.756,0.000){2}{\usebox{\plotpoint}}
\put(945.55,530.86){\usebox{\plotpoint}}
\multiput(965,534)(20.473,3.412){2}{\usebox{\plotpoint}}
\put(1007.15,539.00){\usebox{\plotpoint}}
\multiput(1026,539)(20.473,3.412){2}{\usebox{\plotpoint}}
\put(1068.57,547.35){\usebox{\plotpoint}}
\multiput(1086,552)(20.097,5.186){2}{\usebox{\plotpoint}}
\put(1129.10,561.61){\usebox{\plotpoint}}
\multiput(1147,564)(19.753,6.372){2}{\usebox{\plotpoint}}
\put(1189.15,577.34){\usebox{\plotpoint}}
\multiput(1208,583)(19.487,7.145){2}{\usebox{\plotpoint}}
\multiput(1238,594)(18.444,9.519){2}{\usebox{\plotpoint}}
\put(1284.83,616.86){\usebox{\plotpoint}}
\multiput(1299,623)(16.669,12.367){2}{\usebox{\plotpoint}}
\multiput(1330,646)(15.173,14.162){2}{\usebox{\plotpoint}}
\multiput(1360,674)(14.676,14.676){2}{\usebox{\plotpoint}}
\put(1390,704){\usebox{\plotpoint}}
\end{picture}

\vskip 2 true cm

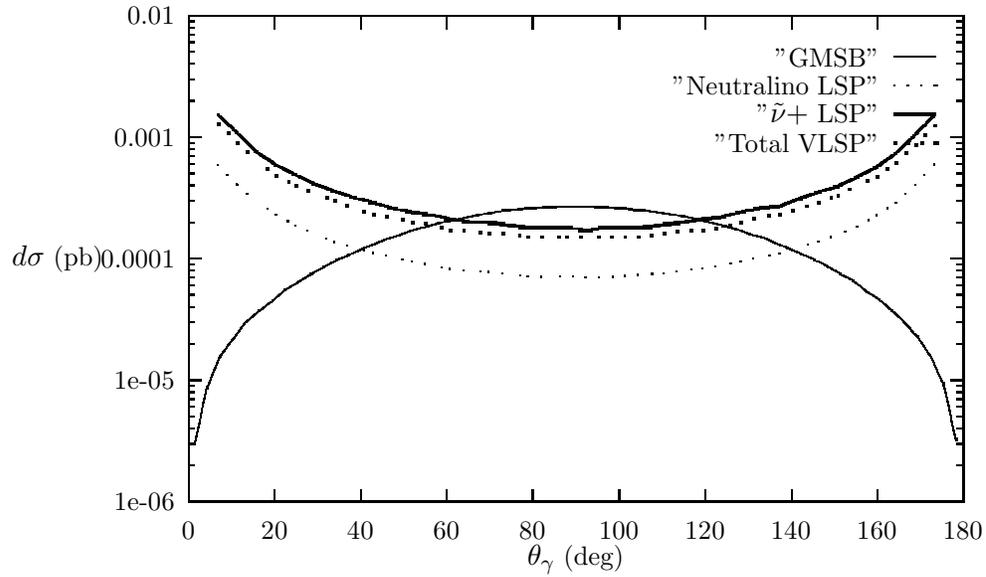
\centerline {\bf Fig. 6}

\noindent Angular distribution of the emitted photon with energy cuts as stated
in the text. The parameter space for GMSB is the same as in Figure 1, 
while for MSSM, the three curves correspond to the parameter sets shown in
Figure 4. 

\end{document}